\documentclass[a4paper,12pt]{article} 

\usepackage[sort&compress,numbers]{natbib}

\usepackage{makecell}
\usepackage{adjustbox}
\usepackage{fancyhdr}
\usepackage{url}
\usepackage{xspace}
\usepackage{numprint}
\usepackage{soul}
\usepackage{graphicx}

\usepackage{tabularx}

\usepackage{amsmath}
\usepackage{amssymb}
\usepackage{amsthm}

\newcommand{\indeg}{\deg^-}

\newcommand{\dist}[2]{\operatorname{dist}(#1,#2)}

\newcommand{\eg}{e.\,g.,\xspace}
\newcommand{\ie}{i.\,e.,\xspace}
\newcommand{\etal}{\textit{et al.}\xspace}

\newcommand{\Guu}{G^{uu}}
\newcommand{\Guw}{G^{uw}}
\newcommand{\Gdu}{G^{du}}
\newcommand{\Gdw}{G^{dw}}

\newcommand{\COD}{C_{OD}}
\newcommand{\CID}{C_{ID}}
\newcommand{\COS}{C_{OS}}
\newcommand{\CIS}{C_{IS}}
\newcommand{\CBuu}{C_{B}^{uu}}
\newcommand{\CBuw}{C_{B}^{uw}}
\newcommand{\CCdu}{C_{C}^{du}}
\newcommand{\CCdw}{C_{C}^{dw}}
\newcommand{\CtCdw}{\tilde{C}_C^{dw}}
\newcommand{\CEuu}{C_{E}^{uu}}
\newcommand{\CKatzdu}{C_{Katz}^{du}}
\newcommand{\CKatzdwrightarrow}{C_{Katz\rightarrow}^{dw}}
\newcommand{\CGCw}{C_{GC}^{w}}
\newcommand{\rhn}[1]{N_r(#1)}
\newcommand{\SCfa}{SC_{1}}
\newcommand{\SKfa}{SK_{1}}
\newcommand{\SKfb}{SK_{2}}
\newcommand{\SKfc}{SK_{3}}

\usepackage[backgroundcolor=yellow]{todonotes} 

\newcommand{\hmeyhidden}[1]{}

\newcommand{\msimhidden}[1]{}

\newcommand{\notehidden}[1]{}

\begin{document}
\setcitestyle{numbers,sort}
\title{Combined Centrality Measures for an Improved Characterization of Influence Spread in Social Networks}

\maketitle
\begin{center}
\author{{
\sc Mehmet Simsek},\\[2pt]
D\"{u}zce University, Faculty of Engineering, Department of Computer Engineering, D\"{u}zce, Turkey\\
{\sc Henning Meyerhenke}\\[2pt]
Humboldt-Universit\"{a}t zu Berlin, Department of Computer Science, Berlin, Germany\\
{meyerhenke@hu-berlin.de}}
\end{center}

\begin{abstract}
{Influence Maximization (IM) aims at finding the most influential users
in a social network, \ie users who maximize the spread of an opinion
within a certain propagation model. 
Previous work investigated the correlation between influence spread
and nodal centrality measures to bypass more expensive IM simulations.
The results were promising but incomplete, since these studies investigated 
the performance (\ie the ability to identify influential users) of centrality measures 
only in restricted settings, \eg in undirected/unweighted networks and/or 
within a propagation model less common for IM.

In this paper, we first show that good results within the
Susceptible-Infected-Removed (SIR) propagation model for unweighted and undirected networks
do not necessarily transfer to directed or weighted networks under the popular Independent Cascade (IC) propagation model.
Then, we identify a set of centrality measures with good performance
for weighted and directed networks within the IC model.
Our main contribution is a new way to combine the centrality measures 
in a closed formula to yield even better results.
Additionally, we also extend gravitational centrality (GC) with the proposed combined centrality measures.
Our experiments on 50 real-world data sets show that our proposed centrality measures outperform well-known centrality measures and the state-of-the art GC measure significantly.}
{social networks, influence maximization, centrality measures, IC propagation model, influential spreaders }
\\
\end{abstract}

\section{Introduction}
\label{sec:intro}
%
\emph{Context--}
%
Online Social Networks (OSNs) are platforms where many people are connected to each other, \eg due to their friendship
or due to sharing similar opinions~\citep{Henri2003,Zareie2018}. In recent years, with the expansion of OSNs, 
modeling and analyzing the spread of an impact on the network (opinion, information, unwanted content, viruses, etc.) 
has gained importance~\citep
{Kimura2010,Sheikhahmadi2017}.
%
Deeper insights into impact propagation and key players in this process can be very beneficial,
\eg by maximizing the spread of an advertisement~\citep
{Chevalier2006,Probst2013} or by preventing the (typically rapid)
spread of a rumor, virus, or epidemic~\citep
{Madar2004,Pastor-Satorras2001}.

Finding the key players is formalized as the \emph{Influence Maximization} (IM) problem, which
asks for the set of $k$ nodes with the highest number of influenced users 
(\ie the \emph{influence spread})~\citep
{Li2018c}. How the influence spreads, is captured by
a so-called \emph{propagation model}, also see Section~\ref{sub:im-social-prop}. It has been shown that the IM problem is $\mathcal{NP}$-hard
under most propagation models~\citep
{Kempe2003}. Thus, when addressing IM in practice, one usually opts for 
heuristic approaches or even proxies such as centrality measures.
Centrality measures indicate the importance of a node in the network via its position~\citep
{Borgatti2006b,Borgatti2005};
numerical values yield a partial order and thus a node ranking. 
Such a ranking is an important basis for seeding the key players in many IM algorithms~\citep
{Kempe2003,Borgatti2006,Borgatti2009}.
Also, numerous recent works investigated the correlation between centrality values of nodes and their influence 
capability~\citep
{Zareie2018,Ma2016,Wang2018,Namtirtha2018,Liu2016b,Berahmand2018,Salavati2018}
-- not only for established measures such as betweenness, closeness or Katz centrality, but also
for newly developed centrality measures such as Gravitational Centrality (GC)~\citep
{Ma2016}.
In this case the centrality measures act as a proxy, \ie they indicate the influence capability of a node implicitly.
With a good correlation, one may be able to bypass more costly propagation simulations.\\ \\
\emph{Motivation--} The propagation model is indeed an integral part of the IM problem which determines the key players -- 
thus, different models may lead to a completely different set of influencers. 
A centrality measure's ability to indicate the influence spread capability of a node (\ie its \emph{performance}
in our context) is affected by the propagation model as well. 
Indeed, a centrality measure that provides good performance on undirected and unweighted networks under 
the Susceptible-Infected-Removed (SIR) propagation model~\citep
{Newman2018networks}, may give poor results 
on directed and weighted networks under the Independent Cascade (IC) model. 
Most of the established and recently tailored centrality measures, however, have been investigated under the SIR model and similar models such as Susceptible-Infected (SI) only~\citep
{Zareie2018,Ma2016,Wang2018,Namtirtha2018,Liu2016b,Salavati2018,Li2016b}. Most of the recent IM algorithms,in turn, have been developed for Independent Cascade \citep
{Li2014,Gong2016,Simsek2018,Yang2018,Li2017d,Gong2016a} and partly for
Linear Threshold~\citep
{Song2015,Chen2009}.
Hence, in this study, we focus on IC propagation and aim at centrality measures that 
correlate well with the nodes' influence capabilities under IC. \\ \\
\emph{Contribution and Outline of the Paper--}
To this end, after preprocessing~(Section~\ref{sub:preprocessing}), we analyze numerous centrality measures on 50 real-world data sets under the IC model, see Section~\ref{sub:exploration}.
Their performance in terms of correlation to influence spread often differs significantly from their
performance in the SIR model. For example, GC's performance is much worse with IC.
This is an important observation since most of the recent centrality measure development studies have been tested under (and partially tuned for) the SIR model~\citep
{Zareie2018,Ma2016,Wang2018,Namtirtha2018,Liu2016b,Salavati2018,Li2016b}.
To the best of our knowledge, our study is the most comprehensive one on new centrality measures for the IC model.

We put the best performing centrality measures together as linear combinations of
two each; this yields four new combined centrality measures (Section~\ref{sub:combined-centralities}).
To obtain the coefficients of each single measure, we use the correlation between the centrality
measure and the real spreading capability.

In addition, we develop new measures based on Gravitational Centrality; instead of the original $k$-shell
mass (see Section~\ref{sub:gc-etc} for the definition), we use our combined measures.

Our experimental results (cf.\ Section~\ref{sec:exp-results}) show that the proposed combined centrality
measures and the modified GC measures outperform the state of the art significantly (the latter being based 
on GC and some basic centrality measures). 
Thus, with the proposed measures, one can bypass costlier propagation simulations in the IC model,
but still gets highly correlated results.

\section{Preliminaries and Related Work}
\label{sec:prelim-related}
\subsection{Notation}
\label{sub:notation}
We represent a social network by a weighted simple graph $G=\{V,E, w\}$\footnote{We thus use the terms \emph{network} and \emph{graph} interchangeably in this paper.},
which is directed unless stated otherwise.
$V$ is the set of nodes (individuals), $E$ is the set of edges (relations), 
and $w: E \rightarrow \mathbb{R}_{>0}$ is the edge weight function.

In our context we usually encounter weighted graphs where the edge weights model the
influence diffusion probabilities between neighboring nodes.
For an easier distinction, we write $\Guu$ for undirected unweighted graphs,
$\Guw$ for undirected weighted graphs,
$\Gdu$ for directed unweighted graphs, and
$\Gdw$ for directed weighted graphs, respectively.

We frequently use the (weighted) adjacency matrix $A$ of $G$, which contains the weight
of edge $(u,v)$ in position $(A)_{u,v}$ (often written as $a_{uv}$) and zeros elsewhere.
Finally, the \emph{$r$-hop neighborhood}, $\rhn{u}$, of a node $u$ is the set of nodes that can be
reached from $u$ by traversing at most $r$ edges.

\subsection{Influence Maximization in Social Networks}
\label{sub:im-social}
Influence maximization (IM) aims at finding a small subset of nodes that are able to influence as many
other nodes as possible in a network~\citep{Kempe2003}.
In this context we mean by ``$u$ influences $v$'' if $u$ passes an opinion/information
on to $v$ (possibly indirectly via other nodes) that is accepted by $v$ (and then passed on).
There are many algorithmic approaches to address
this $\mathcal{NP}$-hard combinatorial optimization problem by selecting (hopefully) very influential
seed nodes: various greedy approaches (one-by-one~\citep
{Kempe2003}, single stage seeding, sequential seeding~\citep
{Liu2018}) as well as metaheuristics such as genetic algorithms, simulated annealing, and swarm intelligence~\citep
{Simsek2018,Yang2018,Nunez-Gonzalez2016,Li2017d,Tong2017}.

Regardless of the adopted algorithmic approach, it is rather natural to evaluate the nodes in terms of their
influence spread capability -- numerical values for this evaluation can then lead to a ranking.
Typically, such an evaluation is based on propagation simulations, which are very costly.
The results of these simulations depend very much on the propagation model,
\ie how an opinion is passed on (or not) to the neighbors of a node.
The two models most relevant for our paper are described next.

\subsection{Propagation Models for IM in Social Networks}
\label{sub:im-social-prop}
Propagation (or diffusion) models can be categorized into three main types:
(i) Threshold models such as Linear threshold (LT)~\citep
{Peng2017,Tong2016},
(ii) cascading models such as Independent cascade (IC)~\citep
{Song2015}, and 
(iii) epidemic models such as Susceptible-Infected-Removed (SIR)~\citep
{Salavati2018}.
This paper focuses on IC; since IC can be seen as a variant of SIR, we describe both
in some detail and pass over LT.

The SIR model is a general information diffusion model often used in modeling disease spread; 
each node has three states: susceptible (S), infected (I), and recovered (R). 
Infections can only happen when an infected node transmits the disease to a neighboring susceptible node.
In each discrete time step, the infected nodes can spread the disease with  probability $\beta$, 
then enter the recovered state with another probability.
In the context of IM, the information to be spread is the disease in SIR, of course.
The original and frequently used SIR model (see for example~\citep
{Newman2018networks,Kamp2013a})
does not reflect the behavior of influence spread in OSNs
since it assumes one global infection probability, regardless of the node pair involved
(although more general SIR variations exist~\citep
{Sun2014a,Tolic2018a}, but not in the IM context). 

The IC model, our main focus, can be considered as a close relative SIR, though.
IC has only two states (active and inactive and thus no recovered state), but also a static
(\ie unchanged over time) $\beta$ value.
If a person is influenced by another person, it becomes active. An activated person can influence other persons and 
cannot return to the inactive state again. 
The IC model associates with each $e \in E$ a propagation probability $P(e) \in [0,1]$.
If we already know sensible values for these probabilities of influence diffusion, then we can use this information.
However, if they are unknown, the literature usually resorts to established probability models.
So do we: we adopt the Weighted Cascade Setting (WCS) model in which for $e = (u, v) \in E$ one sets $P(e) = 1/(\indeg(v))$, where $\indeg(\cdot)$ is the in-degree.
WCS is based on the idea that a nodeâs probability of being influenced is inversely proportional to the number of nodes that may directly influence this node. 

Influence diffusion in the IC model works as follows: A set of initially active (= influenced) nodes, the \emph{seed} nodes, is chosen. Then, within each iteration, all active nodes try to influence all their out-neighbors. 
To this end, each active node generates a random number $r_e \in [0,1]$ per out-edge $e$. If $r_e < P(e)$, 

then the neighbor at the other end of $e$ is activated. If no new node is activated in an iteration, the propagation process
ends. Since the IC model is probabilistic, modeling the propagation needs to be repeated and the expected value of the 
propagation should be taken. It is usually enough (from an empirical point of view) to repeat the 
propagation \numprint{20000} times~\citep
{Tong2016}.

\subsection{Established Centrality Measures}
Recall that centrality measures are used to rank nodes based on their position in the graph. 
This ranking can also be used to seed IM algorithms with very central nodes~\citep
{Kimura2010, Ma2016,Wang2018,Namtirtha2018,Liu2016b,Berahmand2018,Salavati2018,Tong2016,Liu2016}.
Since such a ranking is often much faster to compute, centrality measures have been of high interest
in the context of IM. We are interested in measures with high \emph{performance},
\ie whose ranking result correlates well with the influence spread of the nodes.
After describing established basic measures (see \eg Newman~\citep
{Newman2018networks}) first,
we review measures created with IM in mind.

Basic local centrality measures are \emph{degree} and \emph{strength centrality}.
The degree centrality of $u \in V$, $C_D(u)$, is the size of $u$'s neighborhood,
\ie the number of $u$'s neighbors. This can be generalized in an analogous manner to in- and out-degree
centrality ($\CID(u)$ and $\COD(u)$), respectively, in directed graphs.
Strength centrality, $C_S(u)$, is just the weighted version of degree centrality: instead of using
the neighborhood \emph{size}, one sums up the weight of all incident edges of $u$.
As above, this notion can be generalized easily to in- and out-strength centrality
($\CIS(u)$ and $\COS(u)$) in directed graphs, respectively.

One of the global centrality measures is betweenness centrality. It considers a node's 
participation in shortest paths:
\begin{equation} \label{eq:3}
C_B (u) = {\sum_{s\neq u \neq t \in V}\frac{\sigma_{st}(u)}{\sigma_{st}}},
\end{equation}
where $\sigma_{st}$ is the number of all shortest paths between/from $s$ and/to $t$ and $\sigma_{st}(u)$  is the number of shortest paths between/from $s$ and/to $t$ that pass through $u$ as intermediate node.
If a node's betweenness is high, more information is assumed to flow through this node.

Also based on shortest paths is the global measure closeness centrality; 
it is defined as the reciprocal of the average distance $\operatorname{dist}(\cdot)$ 
(distance = length of shortest path) to all other nodes.
This way, a high closeness value indicates that the corresponding node is located in the center
of the graph:
\begin{equation} 
\label{eq:4}
C_C (u) = \frac{n-1}{\sum_{v \neq u \in V} \dist{u}{v} }.
\end{equation}

Another global measure is eigenvector centrality. It measures a node's importance by the importance of its neighbors.
More precisely, the centrality value of node $u$ is the
$u$th entry of the leading eigenvector $x$ of the adjacency matrix $A$~\citep
{Newman2018networks}. Hence:
%
$
x_u={\lambda_1^{-1} \sum_{v=1}^{n}a_{uv}x_v},
$
where $\lambda_1$ is the largest eigenvalue of $A$. Eigenvector centrality should not be applied
to directed graphs that are not strongly connected.

We mention Katz centrality as the last global centrality measure:

\begin{equation}
\label{eq:6}
C_{Katz} (u)={\sum_{k=1}^{\infty} \alpha^k \sum_{v=1}^{n}(A^k)_{vu}}.
\end{equation}
Here, $0 < \alpha < 1$ is an attenuation factor to dampen the contribution of the number of
walks of length $k$ from $v$ to $u$, $(A^k)_{vu}$, for larger $k$.
\subsection{Recent Centrality Measures for IM Seeding}
\label{sub:gc-etc}
Additionally, several centrality measures have been developed for or adap\-ted to certain IM
propagation models recently -- 
for example gravitational centrality (GC)~\citep
{Ma2016}, $C_{GC}$, which is inspired by Newton's
gravity formula:
\begin{equation} \label{eq:7}
C_{GC} (u)={\sum_{v\in N_{r}(u)}\frac{ks(u)ks(v)}{(\dist{u}{v})^2}}
\end{equation}
Here, $ks(u)$ and $ks(v)$ are the \emph{k-shell values} of nodes $u$ and $v$, respectively.
The k-shell of a graph $G$ is a subgraph that consists of the nodes in the $k$-core but not in the $(k+1)$-core.
The $k$-core of $G$, in turn, is the maximal subgraph in which every node has degree at least $k$~\citep
{Newman2018networks}. 
As the original paper~\citep
{Wang2018}, we set $r := 3$ for the neighborhood $N_r(\cdot)$
in all GC-related measures. 

Ma \etal~\citep
{Ma2016} use the SIR epidemic model to investigate the performance of gravitational centrality and an extension of GC, GC+, for IM.
They compare its IM performance with established centrality measures such as degree, closeness, betweenness, semi-local\footnote{Semi-local centrality extends degree centrality by not only considering direct neighbors, 
but also two-hop neighbors~\citep
{Chen2012a}.} centrality, etc. 
The average performance of the two new measures in terms of ranking correlation and distinction between nodes
is slightly better than for established measures;
in terms of Kendall $\tau$ correlation (defined in~\ref{sec:perf-measures}) results aggregated over nine
real-world data sets, the performance of GC and GC+ are $0.83$ and $0.847$, respectively. The performance of 
semi-local centrality, the best competitor in the study, is $0.821$.

Originally, GC has been developed for undirected and unweighted graphs,
but it can be generalized for weighted networks as well~\citep
{Ma2016}.
To this end, a partially weighted degree needs to be calculated:
\begin{equation} \label{eq:8}
k_i'={\sqrt{k_i\sum_{j=1}^{n}a_{ij}}}
\end{equation}
Here, $k_i$ is the (unweighted) degree of node $i$, so that we take the square root
of the product of the unweighted and the weighted degree.
Garas \etal~\citep
{Garas2012} normalize the $k_i'$ values. As we work with directed
graphs in which some nodes have out-strength $0$, we adapt their normalization process and do
not divide by the minimum value. The experimental results and their interpretation remain
unaffected by this change.

From now on, we refer to gravitational centrality for weighted networks as $\CGCw$.

Wang \etal~\citep
{Wang2018} recently proposed two extensions of gravitational centrality,
$ks_G$ and $ks_{G+}$. 

$ks_G$ modifies the $C_{GC}$ formula by using degree values instead of $ks(v)$ in Eq.~(\ref{eq:7}). $ks_{G+}$ is calculated as sum of all neighbors' $ks_G$ values. The experimental results reveal that the performance of
$ks_G$ and $ks_{G+}$ is similar to that of $C_{GC}$ in the SIR model.

Other recent developments in the field include BridgeRank~\citep
{Salavati2018} and dyna\-mics-sensitive 
(DS) centrality~\citep
{Liu2016b}. BridgeRank is a semi-local measure based on communities and
local betweenness values. SIR experiments on four real-world and four synthetic networks~\citep
{Liu2016b} have shown that BridgeRank and its variants outperform basic centrality measures.
DS, in turn, integrates topological features of the network and the spreading dynamics of the propagation model under consideration. SIR and SI experiments on four real-world networks indicate that DS can outperform basic centrality measures as well.

GC is a rather general framework; different centrality measures can be included,
in particular in the numerator. New studies are inspired by GC, \eg 
a new $k$-shell hybrid method has been developed for unweighted networks
and the SIR model~\citep
{Namtirtha2018}. Their results, if compared to GC,
do not significantly outperform the original definition~\citep
{Salavati2018,Liu2016b,Namtirtha2018}, though.
Moreover, all studies mentioned in this papragraph work with unweighted and/or undirected networks. Thus, one can still consider GC as state of the art in the field.
While an extension to weighted GC has been proposed~\citep
{Ma2016}, to the best of
our knowledge we provide the first substantial experimental results for its usage in weighted networks.

\subsection{Summary}
Greedy algorithms as well as metaheuristics compute their seed set based on the fitness of the nodes --
in this case the real, simulated or indirectly assumed influence spread capabilities. Using centrality measures
as a proxy for that can save a lot of running time -- if the correlation between centrality and influence
spread is high. 
As reported in the literature review, most recent works investigated centrality measures under the SIR model
and very similar models. Yet, most of the recent IM algorithms have been developed for the 
Independent Cascade (IC) model (and partly for Linear Threshold).
As the performance of a centrality measure depends on the propagation model (among others),
good performance on undirected and unweighted networks under the 
SIR propagation model~\citep
{Ma2016} 
may not transfer to directed and weighted networks under the IC model.
Hence, we aim at new centrality measures with good performance
under the (for IM) more popular IC model.

\section{Materials and Methods}
\label{sec:materials-methods}
Our primary assumption is that a high correlation between the centrality scores of
one measure and the influence spread can be further improved by combining two
measures appropriately. Thus, in this section, we identify centrality measure
candidates by exploration and derive new measures, \eg as linear combinations of two candidates.

\subsection{Data Acquisition and Preprocessing}
\label{sub:preprocessing}
The 50 social networks we use for our study are listed in Table~\ref{table:input-data} in~\ref{sec:input-data}.
They have been downloaded from the public sources SNAP~\citep
{snapnets}, KONECT~\citep
{Kunegis2013}, and 
Network Repository~\citep
{nr-aaai15}. We can use each network in principle in four ways:
The undirected networks are made directed by pointing
each edge from the node with smaller to the node with higher ID. When considering undirected graphs only,
we ignore the direction specified by directed graphs. Similarly, when considering unweighted graphs only,
we ignore weights specified by the data sets. On the other hand, when we work with a collection of
weighted graphs, we create weighted versions of unweighted data sets by using the WCS model (see Section~\ref{sub:im-social-prop} for WCS).

This leads to four data collections that we index according to their type: (i) $\Guu$: undirected and unweighted, (ii) $\Guw$: undirected and weighted, (iii) $\Gdu$: directed and unweighted, (iv) $\Gdw$: directed and weighted. We distinguish between four types because not every centrality measure can be computed
for all types (without problems). Our main focus is on $\Gdw$, however, because it is 
the most relevant input class for IM in social networks.

In the IC model, a high edge weight means a high probability of a node $u$ to be influenced by another node $v$.
For closeness and betweenness, however, it means that two nodes are further away.
Thus, we invert the edge weights of the graphs as $1/w$ when calculating distance-based centrality measures such as $\CBuw$ and $\CCdw$.

\subsection{Exploratory Experiments}
\label{sub:exploration}
We start by investigating the correlation of single centrality measures and their influence spread.
Recall that our plan is to combine the successful ones later on to obtain an even better measure.
For all our experiments in this paper, we use NetworKit~\citep
{STAUDT2016} as network analysis tool.
Self-implemented code for the new measures is written in Python~3.

We investigate the following centrality measures on the different graph types as specified:
(i) Out-degree in $\Gdu$ ($\COD$), (ii) out-strength in $\Gdw$ ($\COS$), (iii) betweenness in $\Guu$
($\CBuu$), (iv) betweenness in $\Guw$ ($\CBuw$), (v) outbound closeness in $\Gdu$ ($\CCdu$), (vi) outbound
closeness in $\Gdw$ ($\CCdw$), (vi) eigenvector centrality in $\Guu$ ($\CEuu$), (vii) Katz (incoming) centrality
in $\Gdu$ ($\CKatzdu$), and (viii) Katz outgoing centrality in $\Gdw$ ($\CKatzdwrightarrow$).
\label{Modified_Katz}
Moreover, since exact betweenness calculations are very time-consuming (even more than simulating influence propagation),
we resort to approximations based on the algorithms KADABRA~\citep
{Borassi2019} for undirected unweighted graphs 
and RK for undirected weighted graphs~\citep
{DBLP:journals/datamine/RiondatoK16} (with $\epsilon = 0.1$).
For closeness centrality, we use the corresponding approximation algorithm (with $\epsilon = 0.1$)
in NetworKit, too.
Also note that Katz centrality takes incoming edges into account.
Yet, on the social networks we use, the direction of the influence
is modeled by outgoing edges. That is why we can expect low
Katz centrality to co-occur with high influence spread.

The expected influence spread is calculated for the IC model as follows:
Each node is selected as single seed, then the propagation based on this seed is run until convergence.
This probabilistic process is repeated \numprint{20000} times (as suggested by~\citep
{Tong2016}) per node to account for random
fluctuations; the arithmetic mean is used as the influence spread result for that seed node.

\begin{figure}[tb]
  \includegraphics[width=0.94\linewidth]{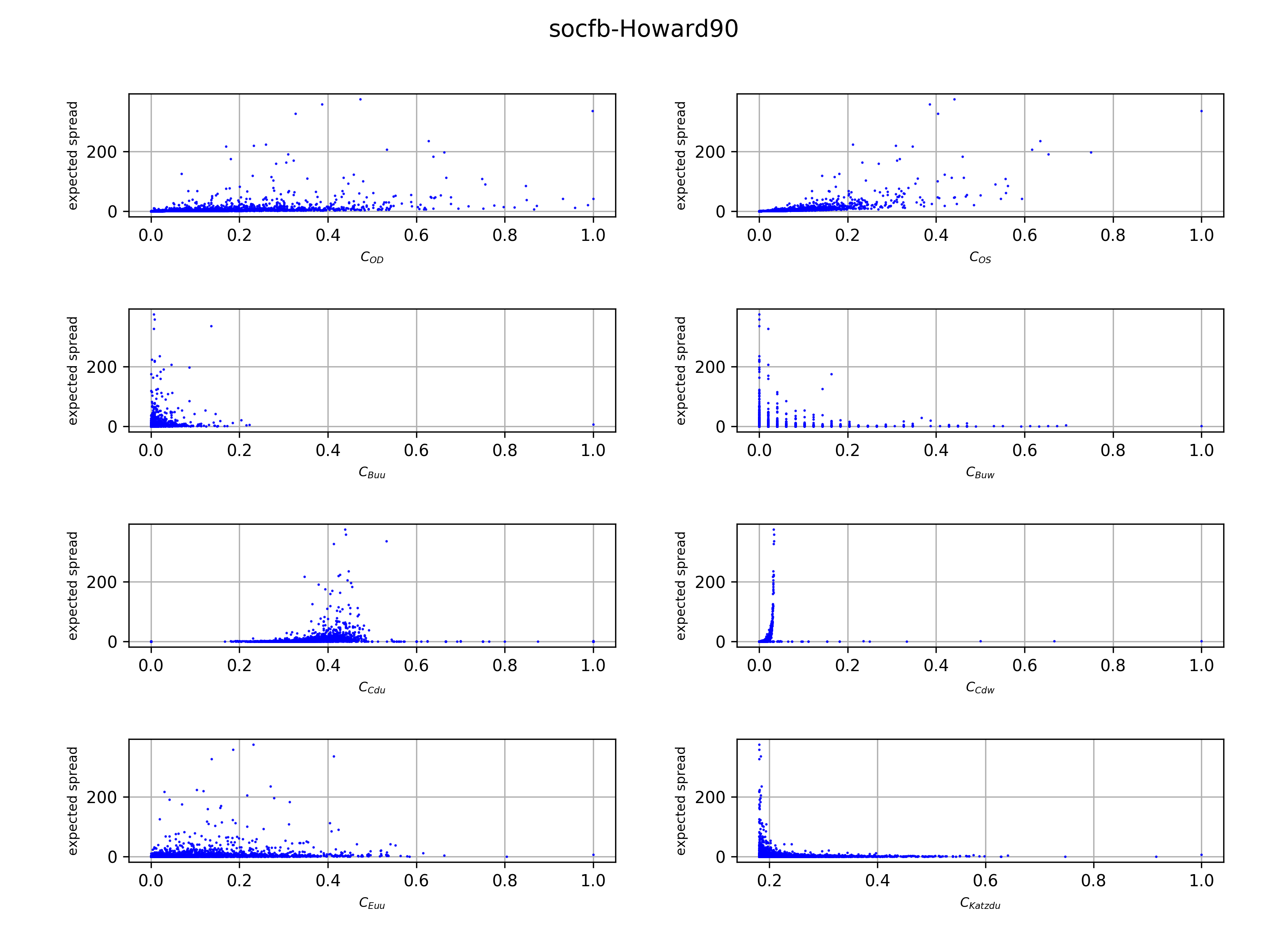}
  \caption{Centrality measure vs. expected spread for the socfb-Howard90 data set.}
  \label{fig:fig1}
\end{figure}

Figure~\ref{fig:fig1} displays the simulation results for the network 
socfb-Howard90.\footnote{This network has been selected as its results
are representative of the results for the data collection at large.
}
Each point $(x,y)$ represents the centrality score $x$ (usually normalized by the maximum value) vs.\ its influence spread $y$ for a particular seed.
Our visual interpretation is mostly interested in a good distinction between the
nodes and a (possibly linear) trend/correlation between centrality and spread.

We observe that the point distributions of $\COS$, $\CCdw$, and $\CKatzdu$ 
are reasonably spread out. This means that these centrality measures
allow a better distinction between nodes in terms of their influence
spread capability. $\CBuw$, on the other hand, yields values that seem
discretized into numerous narrow intervals. These observations provide
visual (and thus informal) indication of good and bad ranking monotonicity, respectively; in general terms, 
ranking monotonicity measures the fraction of (non-)ties in a ranking and thus
the ability to distinguish nodes (see Def.~\ref{def:monotonicity} in~\ref{sec:perf-measures} for a formal definition).

Furthermore, the $\COS$ values increase with the expected spread.
While this behavior is not necessarily linear, a trend in the sense of
``higher centrality means higher spread'' is visible.
$\COD$ yields results similar to $\COS$, but the $\COS$ values are more distinguishable. 
In fact, this trend is a visual indication of a low ranking error, which
measures (broadly speaking) how well a ranking $X$ preserves a ranking $Y$ 
(a more formal description is given as Definition~\ref{def:kendall-ranking-correlation}
in~\ref{sec:perf-measures}).
Betweenness, in turn, does not have such a trend; its values are clustered
within narrow intervals. 
As expected, the slope of the $\CKatzdu$ values is negative; we will account for that later on.
To analyze the results not only visually, we proceed with the Kendall
$\tau$ ranking correlation coefficient (see Definition~\ref{def:kendall-ranking-correlation} in~\ref{sec:perf-measures}) for each measure. The numerical results are given in 
Table~\ref{table:kendall-explore} in~\ref{sec:app-exp-data}.
In order to see a measure's relative performance in comparison with $\COD$, we divide
all scores by the $\COD$ results. For a global, aggregate perspective, we
provide in the last row the geometric means of these ratios over all data sets.
(We use the absolute values of the results in the geometric mean calculation because some results are negative. While this approach may make the means harder to
interpret, a close inspection of Table~\ref{table:kendall-explore} reveals that
the qualitative interpretation is not changed.)
It is evident from the values that the centrality measures betweenness, 
closeness, eigenvector, Katz, and wks do not perform well when used alone. 
Only strength centrality shows a high correlation. 
Of the global measures, closeness and Katz perform best.

In summary, the best performing local measure is $\COS$.
Of the global measures, $\CCdw$ and $\CKatzdu$ seem overall reasonably promising.
The other measures' patterns are not as good or even unclear. 
Therefore, we proceed with $\COS$, $\CCdw$, and $\CKatzdu$ when 
creating combined measures; further experiments in the paper will not consider the
other basic measures.

\subsection{Combined Centrality Measures}
\label{sub:combined-centralities}
Based on the insights above, we proceed by combining different measures 
into new ones. We aim for a linear combination of a local and a global
centrality measure to merge both perspectives of a node,
\eg $\SCfa := \gamma \cdot \COS + \delta \cdot \CtCdw$.
Note that we introduce a small change to the closeness values in weighted directed graphs. 
Since the largest spread is observed for $\CCdw$ values between $0.0$ and $0.04$ 
(with high slope), we modify $\CCdw$ in order to match high centrality values with high influence spread:
\begin{equation} 
\label{eq:10}
\CtCdw(u) := 
\begin{cases}
    \CCdw(u) + 1 - 0.04 & \text{if } \CCdw(u) \leq 0.04 \\
    1 - \CCdw(u) + 0.04 & \text{if } \CCdw(u) > 0.04
\end{cases}
\end{equation}

To determine the coefficients $\gamma$ and $\delta$ of our linear
combination, we make use of the
respective correlation strength -- the measure with higher strength shall receive a higher coefficient.
To this end, let $k(\COS)$ and $k(\CtCdw)$ be the geometric mean of the normalized $\tau$ results of $\COS$ and $\CtCdw$
(see Table~\ref{table:kendall-explore}). This leads to
$\gamma := \frac{k(\COS)}{k(\COS)+k(\CtCdw)} \approx 0.64$ and $\delta := \frac{k(\CtCdw)}{k(\COS)+k(\CtCdw)} \approx 0.36$. Thus:
\begin{equation} \label{eq:9}
\SCfa={0.64 \cdot \COS + 0.36 \cdot \CtCdw}
\end{equation}

As further combinations of a local ($\COS$) and a global (now $\CKatzdu$) measure, we propose the following group of measures:
\begin{equation} \label{eq:11}
\SKfa=\COS+{\frac{\COS}{\CKatzdu}}
\end{equation} \par
\begin{equation} \label{eq:12}
\SKfb=\COS-{\frac{\CKatzdu}{\COS+\CKatzdu}}
\end{equation} \par
\begin{equation} \label{eq:13}
\SKfc={\frac{\COS}{\CKatzdu}}
\end{equation}
The rationale behind these formulas is the following:
In all data sets, the $\COS$ values increase while the expected spread
increases; also, in most of the data sets, the $\CKatzdu$ values 
decrease while the expected spread increases. Thus, we use $\COS$ as
numerator and $\CKatzdu$ as denominator in $\SKfa$ and $\SKfc$.
In $\SKfb$ we use $\CKatzdu$ as numerator because we subtract the 
result of the division from $\COS$ to get a positive correlation
between $\SKfb$ and the expected spread.

\subsection{Variations of Gravitational Centrality}
\label{sub:mgc-variants}
In addition to the proposed measures above, we create modifications of gravitational centrality (GC). GC is a general framework in which new measures can be integrated easily.
To do so, we consider the Kendall $\tau$ correlation results of all the measures  presented in Tables~\ref{table:kendall-explore} and~\ref{table:kendall-combined}. 
The following measures yield significant results and are proposed for directed and weighted graphs. Hence, shortest path calculations are performed on $\Gdw$. 
Also, we invert the edge weights of the graphs as $1/w$ when calculating shortest path lengths for all centrality measures
based on GC (incl.\ $\CGCw$). First, let the modified GC using out-degree
strength be
\begin{equation} \label{eq:14}
MGC_{ODS}(i)={\sum_{j\in N_{3}(i)}\frac{ods(i) \cdot ods(j)}{(\dist{i}{j})^2}},
\end{equation}
where $ods(i)=\COD(i) \cdot \COS(i)$ for a node $i \in V$.
All other parameters have the same meaning as in Eq.~(\ref{eq:7}). Note that $\COD$ and $\COS$ are basic
local measures. Yet, accor\-ding to our results, they are strong indicators for influence capability of 
a node on a directed and weighted graph. 
To (potentially) create an even stronger measure, we propose further variations 
within the GC formula, here specified for $i \in V$:
\begin{equation} \label{eq:15}
MGC_{S}(i)={\sum_{j\in N_{3}(i)}\frac{\COS(i) \cdot \COS(j)}{(\dist{i}{j})^2}},
\end{equation}
\begin{equation} \label{eq:16}
MGC_{SC}(i)={\sum_{j\in N_{3}(i)}\frac{\SCfa(i) \cdot \SCfa(j)}{(\dist{i}{j})^2}},
\end{equation}
\begin{equation} \label{eq:17}
MGC_{SK}(i)={\sum_{j\in N_{3}(i)}\frac{\SKfc(i) \cdot \SKfc(i)}{(\dist{i}{j})^2}},
\end{equation}
\begin{equation} \label{eq:18}
MGC_{wk}(i)={\sum_{j\in N_{3}(i)}\frac{wk(i) \cdot wk(j)}{(\dist{i}{j})^2}},
\end{equation}
where $wk(i)=ods(i) \cdot \CKatzdwrightarrow (i)$.


\section{Experimental Results}
\label{sec:exp-results}
In this section, we provide and discuss the experimental results of
our proposed measures, out-strength, and weighted gravitational centrality ($\CGCw$). Our evaluation is based on the assessment of three ranking
performance measures: Kendall $\tau$ correlation, ranking error, and
ranking monotonicity (for the definitions of these measures, cf.\ \ref{sec:perf-measures}).

\paragraph{Kendall $\tau$ correlation}
Geometric mean values of the normalized Kendall $\tau$ ranking results are shown in Tables~\ref{table:kendall-combined-brief} and~\ref{table:kendall-gc-brief} for all 
proposed centrality measures as well as for $\COS$ and $\CGCw$.
Detailed results on all data sets are shown in Tables~\ref{table:kendall-combined} and~\ref{table:kendall-gc}.

\begin{table}[b]
 \small
\caption{Geometric means of normalized Kendall $\tau$ results for combined measures}
\smallskip
\label{table:kendall-combined-brief}

\resizebox*{0.95\textwidth}{0.06\textheight}{
 \begin{tabular}{||c c c c c c c||} 
 \hline
Centrality Measures & $\COS$ & $\SCfa$ & $\SKfa$ & $\SKfb$ & $\SKfc$ & $\CKatzdwrightarrow$  \\ [0.5ex] 
\hline
Geometric mean & 1.29864 & 1.35576 & 1.04994 & 1.28907 & 1.31493 & 1.30657 \\
 \hline
\end{tabular}
}
\end{table}
\begin{table}[h!]
 \small
\caption{Geometric means of normalized Kendall $\tau$ results for modifications of gravitational centrality}
\smallskip
\label{table:kendall-gc-brief}
\resizebox*{0.95\textwidth}{0.06\textheight}{
 \begin{tabular}{||c c c c c c c||} 
 \hline
Centrality Measures & $\CGCw$ & $MGC_{ODS}$ & $MGC_{S}$ & $MGC_{SC}$ & $MGC_{SK}$ & $MGC_{wk}$  \\ [0.5ex] 
\hline
Geometric mean & 0.70205 & 1.38455 & 1.35513 & 1.29802 & 1.31421 & 1.40721 \\
 \hline
\end{tabular}
}
\end{table}

When inspecting the detailed values, we see that $MGC_{wk}$ has the highest correlation on 43 data sets, $MGC_{ODS}$ has the highest Kendall $\tau$ correlation on six data sets, and $\SCfa$ performs best only once.
If we sort the measures according to the geometric mean results in descending order, the ranking is as follows: 
$MGC_{wk}$, $MGC_{ODS}$, $\SCfa$, $MGC_{S}$, $\SKfc$, $MGC_{SK}$, $\CKatzdwrightarrow$, $\COS$, $MGC_{SC}$, $\SKfb$, $\SKfa$, $\CGCw$. 
According to the overall Kendall $\tau$ correlation results, seven of the proposed measures outperform $\COS$, and all of the proposed measures outperform $\CGCw$.
Note that Ma \etal~\citep
{Ma2016} reported that GC's performance in terms of
Kendall's $\tau$ is slightly better than degree centrality's within
the SIR model. In our experiments within the IC model, however, GC reaches only 70\% of out-degree's
performance.

\paragraph{Ranking error}
For the ranking error experiments, we exclude the data sets Moreno highschool, Moreno dutch college, and Moreno seventh grader from the experiments because they have less than 100 nodes and we focus here on the top-$50$.
Geometric means of the normalized ranking error $\epsilon$ for the top-$50$ nodes 
of each measure are shown in Tables~\ref{table:error-combined-brief} and~\ref{table:error-gc-brief}.
The detailed results on all data sets are shown in Tables~\ref{table:error-combined} and~\ref{table:error-gc}.

\begin{table}[h!]
 \small
\caption{Geometric means of normalized $\epsilon$ for $\COS$, $\CKatzdwrightarrow$, and the proposed combined measures}
\smallskip
\label{table:error-combined-brief}
\resizebox*{\textwidth}{0.06\textheight}{
 \begin{tabular}{||c c c c c c c||}
 \hline
 Data sets & $\COS$ & $\SCfa$ & $\SKfa$ & $\SKfb$ & $\SKfc$ & $\CKatzdwrightarrow$  \\ [0.5ex] 
 \hline
geometric mean & 0.47425 & 0.47453 & 0.42411 & 0.43921 & 0.41501 & 0.47415 \\
 \hline
\end{tabular}
}
\end{table}
\begin{table}[h!]
 \small
\caption{Geometric means of normalized $\epsilon$ for modifications of gravitational centrality.}
\smallskip
\label{table:error-gc-brief}
\resizebox*{\textwidth}{0.06\textheight}{
 \begin{tabular}{||c c c c c c c||} 
 \hline
 Centrality Measures & $\CGCw$ & $MGC_{ODS}$ & $MGC_{S}$ & $MGC_{SC}$ & $MGC_{SK}$ & $MGC_{wk}$  \\ [0.5ex] 
 \hline
geometric mean & 0.16681 & 0.08432 & 0.09198 & 0.09174 & 0.09066 & 0.24280 \\
 \hline
\end{tabular}
}
\end{table}

When inspecting the detailed data, we see that $MGC_{SK}$ has the lowest $\epsilon$ on 20 data sets; $MGC_{ODS}$, in turn, performs best on 18 data sets.
Moreover, $MGC_{S}$ and $MGC_{SC}$ have the lowest $\epsilon$ on 12 data set, respectively, whereas $\CGCw$ performs best on six data sets. 
All the measuresâ $\epsilon$ values for the socfb-nips-ego data set are $1$; this coincides with
the $\COS$ result in terms of $\epsilon$ (because of normalization). We conjecture this behavior to
stem from the network's sparsity (the average out-degree is $1.03$).

If we sort the measures according to geometric mean results in ascending order, the ranking is as follows: $MGC_{ODS}$, $MGC_{SK}$, $MGC_{SC}$, $MGC_{S}$, $\CGCw$, $MGC_{wk}$, $\SKfc$, $\SKfa$, $\SKfb$, $\CKatzdwrightarrow$, $\COS$. According to the overall $\epsilon$ results, four of the proposed measures outperform gravitational centrality $\CGCw$ and nine of the proposed measures outperform out-strength $\COS$.

\paragraph{Ranking monotonicity}
When analyzing the ranking monotonicity of the measures, we should keep in mind
that higher $M(R)$ values are better. If $M(R)$ is $1.0$ for a measure, it means
that the measure perfectly distinguishes all nodes (\ie it assigns all nodes to different ranks).
The other extreme is if $M(R)$ is $0.0$; then the measure cannot distinguish the nodes at all (\ie it assigns all nodes to only one rank).
Geometric means of the (unnormalized) ranking monotonicity values are shown in Tables~\ref{table:monotonicity-combined-brief} and~\ref{table:monotonicity-gc-brief}. 
Detailed results on all data sets are shown in Tables~\ref{table:monotonicity-combined} and~\ref{table:monotonicity-gc}.

\begin{table}[h!]
 \small
\caption{Geometric means of ranking monotonicity values for $\COS$, $\CKatzdwrightarrow$, and the proposed combined measures}
\smallskip
\label{table:monotonicity-combined-brief}
\resizebox*{\textwidth}{0.06\textheight}{
 \begin{tabular}{||c c c c c c c||} 
 \hline
 Centrality Measures & $\COS$ & $\SCfa$ & $\SKfa$ & $\SKfb$ & $\SKfc$ & $\CKatzdwrightarrow$  \\ [0.5ex] 
 \hline
geometric mean & 0.97148 & 0.99585 & 0.99435 & 0.99438 & 0.99438 & 0.99603 \\
\hline
\end{tabular}
}
\end{table}
\begin{table}[h!]
 \small
\caption{Geometric means of ranking monotonicity values for modifications of gravitational centrality.}
\smallskip
\label{table:monotonicity-gc-brief}
\resizebox*{\textwidth}{0.06\textheight}{
 \begin{tabular}{||c c c c c c c||} 
 \hline
Centrality Measures & $\CGCw$ & $MGC_{ODS}$ & $MGC_{S}$ & $MGC_{SC}$ & $MGC_{SK}$ & $MGC_{wk}$  \\ [0.5ex] 
 \hline
geometric mean & 0.21740 & 0.99513 & 0.99508 & 0.99521 & 0.99722 & 0.99628 \\
 \hline
\end{tabular}
}
\end{table}

When counting the instances with highest ranking monotonicity, we get the following:
$\SKfa$, $\SKfb$, and $\SKfc$ perform best on 33 data sets, 
$MGC_{SK}$ on 13 data sets, $MGC_{wk}$ on 11 data sets, $\SCfa$ and $C_{Katzdw\epsilon}$ on five data sets, $MGC_{ODS}$, $MGC_{S}$, $MGC_{SC}$ on four data sets, and finally $\COS$ on two data sets.
If we sort the measures according to geometric mean results in descending order, the ranking becomes:
$MGC_{SK}$, $MGC_{wk}$, $\CKatzdwrightarrow$, $\SCfa$, $MGC_{SC}$, $MGC_{ODS}$, $MGC_{S}$, $\SKfb$, $\SKfc$, $\SKfa$, $\COS$, $\CGCw$. According to the overall ranking monotonicity results, all proposed measures  outperform gravitational centrality $\CGCw$ and strength $\COS$.
We conjecture that $\CGCw$ performs so badly here because our directed graphs contain many nodes
with out-strength $0$, often leading to the same rank for them.

\paragraph{Stability}
\begin{figure}[tb]
  \includegraphics[width=0.95\linewidth]{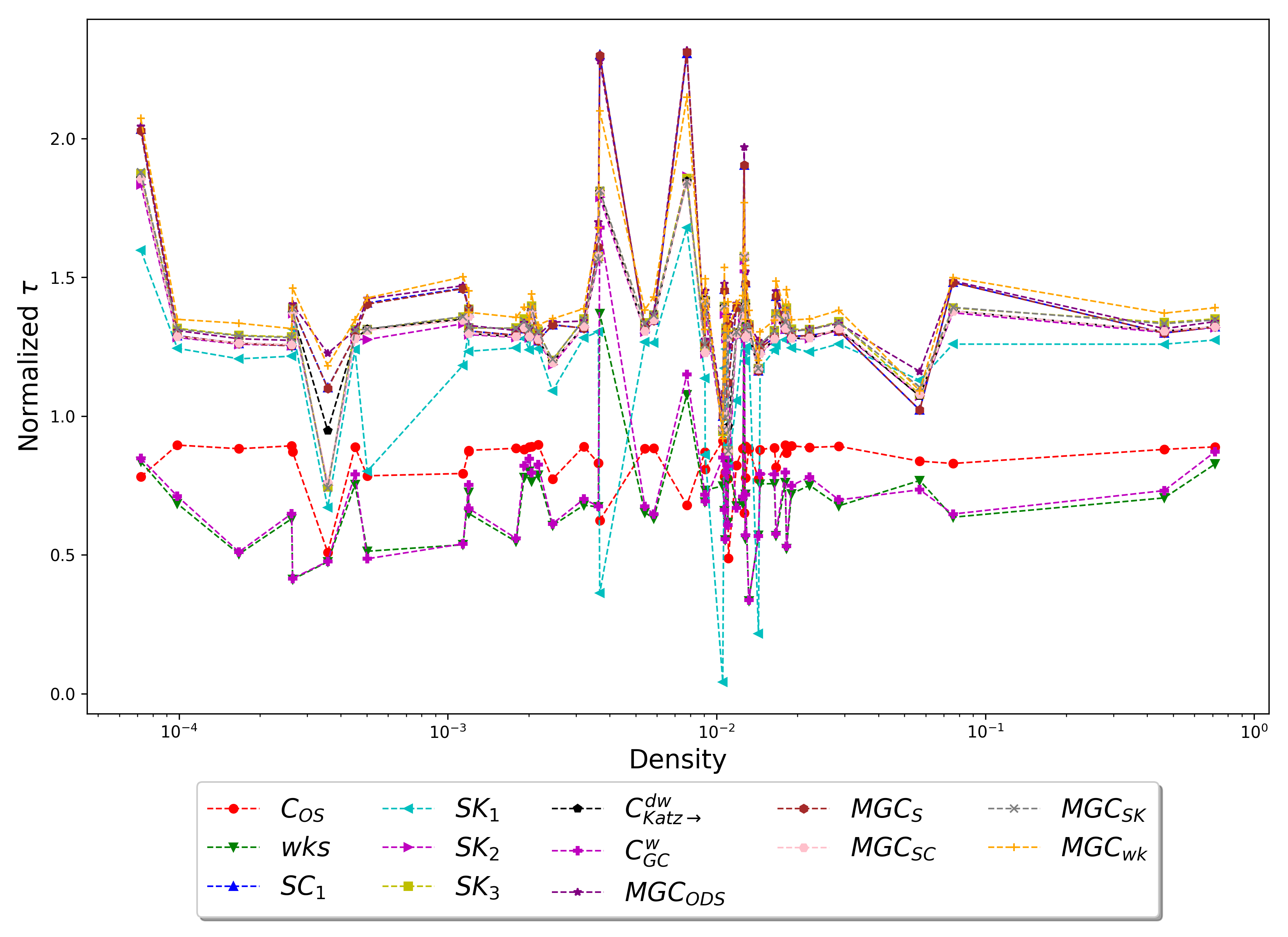}
  \caption{Graph density vs. normalized $\tau$ results of the measures.}
  \label{fig:fig2}
\end{figure}
\begin{figure}[tb]
  \includegraphics[width=0.95\linewidth]{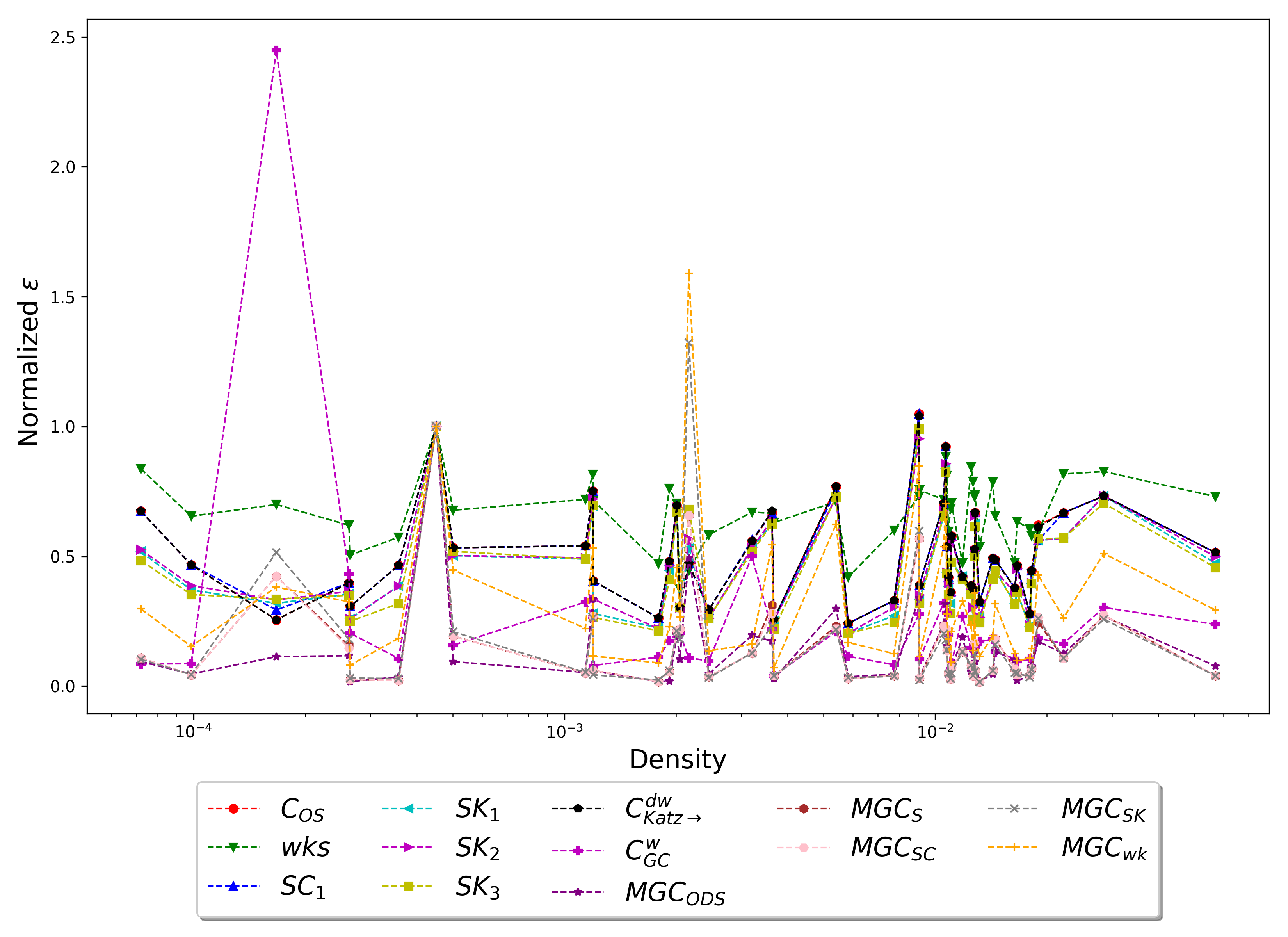}
  \caption{Graph density vs. normalized $\epsilon$ results of the measures.}
  \label{fig:fig3}
\end{figure}
Finally, we inspect graph density vs.\ $\tau$ (Figure~\ref{fig:fig2})
and graph density vs.\ $\epsilon$ (Figure~\ref{fig:fig3}) to assess
the stability of the centrality measures.
In both figures, the $x$-axis corresponds to the graph density of the
data sets, ordered from lowest to highest. 
The data show that the $\tau$ and $\epsilon$ values ($y$-axis) 
of the centrality measures do not change very much with graph density --
but there are some peaks. Thus, the performance of the measures appears
as reasonably stable w.\,r.\,t.\ this parameter.

\paragraph{Running Times}
\label{par:times}
Running times of the proposed combined centrality measures as well as GC and its new variants
were taken on a laptop with Intel Core i5 CPU at 1.6 GHz and 8 GB DDR3 RAM.
Recall that all new measures as well as GC have been implemented in Python; basic centrality measures
(uncombined) can be computed with NetworKit's C++ backend.

In a nutshell, the results are:
$\CGCw$ and its variants fare very similar; also, $\SKfa$'s running time is very close to the running time of $\SKfb$ and $\SKfc$. 
The combined centrality measures require a few seconds at most, often less. 
On average (arithmetic mean), $\SCfa$ takes about one second, $\COS$ and $\SKfa$ only $3$ and $9$ thousandths, respectively.
In contrast, the running time of GC and its variants
can be quite high -- even a few hours for the larger and denser graphs
in our input collection and roughly $28$ minutes on average.
These high running times for GC mostly stem from the iteration over $3$-hop neighborhoods. 
Thus, in this experimental setting, the combined measures are at least three orders of magnitude faster on average.

\section{Conclusion}
\label{sec:conclusions}
We have developed combined centrality measures with high prediction potential for the influence
spread of nodes within the Independent Cascade (IC) model. 
These measures, including their extension of the state-of-the-art measure gravitational centrality (GC),
show a significantly improved empirical performance compared to GC,
both in correlation, monotonicity and in running time.
%
Compared to related studies in the literature,
we add a more meaningful perspective on the
topic by using 50 public real-world data sets as weighted directed
graphs in the common IC propagation model
(as opposed to few unweighted/undirected graphs in the SIR model).
Our new centrality combination with a closed formula takes care to use one local and one global centrality measure together.
This approach unifies the local and global perspective of a node.
Of course, there can be other ways to combine measures,
\eg by using multi-parameter regression.

How to assess the results in terms of the ranking performance measures, depends
very much on the underlying algorithmic approach for IM. If we only pick the
top-$k$ nodes as seeds, then the top-$k$ ranking results with 
lowest error $\epsilon$ are probably most useful (suggesting one of our GC variants).
When applying metaheuristics to IM, Kendall $\tau$ results seems more useful
as it allows a quick evaluation of new seed nodes.
Ranking Monotonicity, on the other hand, is not necessarily useful when inspected in isolation.
But a measure with low ranking monotonicity may be questionable.

\section*{Funding}
This work was supported by The Scientific and Technological Research Council of Turkey (T\"{U}B\.{I}TAK) [Project number: 1059B191700869]; and German Research Foundation (DFG) within Priority Programme 1736 [Grant: ME 3619/3-2].
\section*{Acknowledgement}
This is the preprint of a manuscript that has been published in the Journal of Complex Networks with DOI: 10.1093/comnet/cnz048


\bibliographystyle{unsrtnat}

%


\newpage
\appendix

\section{Input Data}
\label{sec:input-data}
\begin{table}[h!]
\small
\caption{Data set properties}
\label{table:input-data}
\centering
\smallskip
\resizebox*{0.89\textwidth}{0.89 \textheight}{
 \begin{tabular}{||l l l l l||} 
 \hline
 Data sets & $|V|$ & $|E|$ & Average outdegree & Density \\ [0.5ex] 
 \hline\hline
Moreno seventh grader  \citep
{Kunegis2013,Watts1998} & 29 & 376 & 12.97 & 0.4631 \\
Moreno dutch college \citep
{Kunegis2013,VanDeBunt:1999:FNT:593044.593081} & 32 & 710 & 22.19 & 0.7157 \\
Moreno highschool \citep
{Kunegis2013,konect:coleman} & 70 & 366 & 5.23 & 0.0758 \\
Moreno residence hall \citep
{Kunegis2013,Freeman1998} & 217 & 2672 & 12.31 & 0.0570 \\
Moreno  physicians \citep
{Kunegis2013,Coleman1957} & 241 & 1098 & 4.56 & 0.0190 \\
socfb-Haverford76 \citep
{nr-aaai15} & 1446 & 59589 & 41.21 & 0.0285 \\
socfb-Simmons81 \citep
{nr-aaai15} & 1518 & 32988 & 21.73 & 0.0143 \\
socfb-Swarthmore42 \citep
{nr-aaai15} & 1659 & 61050 & 36.80 & 0.0222 \\
Petster hamster friendships \citep
{Kunegis2013} & 1858 & 12534 & 6.75 & 0.0036 \\
Socfb-Amherst41 \citep
{nr-aaai15} & 2235 & 90954 & 40.70 & 0.0182 \\
socfb-Bowdoin47 \citep
{nr-aaai15} & 2252 & 84387 & 37.47 & 0.0166 \\
socfb-Hamilton46 \citep
{nr-aaai15} & 2314 & 96394 & 41.66 & 0.0180 \\
Moreno adolescent health \citep
{Kunegis2013,Moody2001} & 2539 & 12969 & 5.11 & 0.0020 \\
socfb-Trinity100 \citep
{nr-aaai15} & 2613 & 111996 & 42.86 & 0.0164 \\
socfb-USFCA72 \citep
{nr-aaai15} & 2682 & 65252 & 24.33 & 0.0091 \\
socfb-Williams40 \citep
{nr-aaai15} & 2790 & 112986 & 40.50 & 0.0145 \\
socfb-nips-ego \citep
{nr-aaai15} & 2888 & 2981 & 1.03 & 0.0004 \\
socfb-Oberlin44 \citep
{nr-aaai15} & 2920 & 89912 & 30.79 & 0.0105 \\
socfb-Wellesley22 \citep
{nr-aaai15} & 2970 & 94899 & 31.95 & 0.0108 \\
socfb-Smith60 \citep
{nr-aaai15} & 2970 & 97133 & 32.70 & 0.0110 \\
socfb-Vassar85 \citep
{nr-aaai15} & 3068 & 119161 & 38.84 & 0.0127 \\
socfb-Middlebury45 \citep
{nr-aaai15} & 3075 & 124610 & 40.52 & 0.0132 \\
socfb-Pepperdine86 \citep
{nr-aaai15} & 3445 & 152007 & 44.12 & 0.0128 \\
socfb-Colgate88 \citep
{nr-aaai15} & 3482 & 155043 & 44.53 & 0.0128 \\
socfb-Santa74 \citep
{nr-aaai15} & 3578 & 151747 & 42.41 & 0.0119 \\
socfb-Wesleyan43 \citep
{nr-aaai15} & 3593 & 138035 & 38.42 & 0.0107 \\
socfb-Mich67 \citep
{nr-aaai15} & 3748 & 81903 & 21.85 & 0.0058 \\
socfb-Bucknell39 \citep
{nr-aaai15} & 3826 & 158864 & 41.52 & 0.0109 \\
Facebook 2018 tvshow \citep
{Rozemberczki2018} & 3892 & 17262 & 4.44 & 0.0011 \\
socfb-Brandeis99 \citep
{nr-aaai15} & 3898 & 137567 & 35.29 & 0.0091 \\
Facebook combined \citep
{McAuley2012} & 4039 & 88234 & 21.85 & 0.0054 \\
socfb-Howard90 \citep
{nr-aaai15} & 4047 & 204850 & 50.62 & 0.0125 \\
socfb-Rice31 \citep
{nr-aaai15} & 4087 & 184828 & 45.22 & 0.0111 \\
socfb-Rochester38 \citep
{nr-aaai15} & 4563 & 161404 & 35.37 & 0.0078 \\
Facebook 2018 politician \citep
{Rozemberczki2018} & 5908 & 41729 & 7.06 & 0.0012 \\
Advogato \citep
{Kunegis2013,Massa2009} & 6539 & 51127 & 7.82 & 0.0012 \\
Facebook 2018 government \citep
{Rozemberczki2018} & 7057 & 89455 & 12.68 & 0.0018 \\
Wiki-Vote \citep
{Leskovec2010,Leskovec2010a} & 7115 & 103689 & 14.57 & 0.0020 \\
socfb-BC17 \citep
{nr-aaai15} & 11509 & 486967 & 42.31 & 0.0037 \\
Facebook 2018 public figure \citep
{Rozemberczki2018} & 11565 & 67114 & 5.80 & 0.0005 \\
socfb-Columbia2 \citep
{nr-aaai15} & 11770 & 444333 & 37.75 & 0.0032 \\
Facebook 2018 athletes \citep
{Rozemberczki2018} & 13866 & 86858 & 6.26 & 0.0005 \\
socfb-JMU79 \citep
{nr-aaai15} & 14070 & 485564 & 34.51 & 0.0025 \\
Facebook 2018 company \citep
{Rozemberczki2018} & 14113 & 52310 & 3.71 & 0.0003 \\
socfb-UCSB37 \citep
{nr-aaai15} & 14917 & 482215 & 32.33 & 0.0022 \\
socfb-UCF52 \citep
{nr-aaai15} & 14940 & 428989 & 28.71 & 0.0019 \\
Facebook 2018 new sites \citep
{Rozemberczki2018} & 27917 & 206259 & 7.39 & 0.0003 \\
Deezer RO \citep
{Rozemberczki2018} & 41773 & 125826 & 3.01 & 0.0001 \\
Deezer HU \citep
{Rozemberczki2018} & 47538 & 222887 & 4.69 & 0.0001 \\
Deezer HR \citep
{Rozemberczki2018} & 54573 & 498202 & 9.13 & 0.0002 \\
\hline
\end{tabular}
}
\end{table}

\newpage

\section{Performance Measures for Rankings}
\label{sec:perf-measures}
For completeness, we provide here the definitions for known measures for assessing
the performance of rankings.

\newtheorem{definition}{Definition}
\begin{definition}[Kendall $\tau$ ranking correlation~\citep
{Namtirtha2018,KENDALL1945}]
\label{def:kendall-ranking-correlation}
Let $X$ and $Y$ be ranking lists, $n_c$ the number of concordant pairs, and
$n_d$ the number of discordant pairs.
Moreover, let $n_0 := {\binom{n}{2}}$ and $n_1 := \sum_{i} t_i (t_i-1)/2$.
Finally, let $n_2 := \sum_{j} t_j (t_j-1)/2$, where $t_i$ and $t_j$ are the number
of tied values in the $i$th and $j$th group of ties in $X$ and $Y$, respectively.
Then, the \emph{Kendall $\tau$ ranking correlation} between $X$ and $Y$ is:
\begin{equation} \label{eq:19}
\tau(X,Y) := {\frac{n_c - n_d}{\sqrt{(n_0 - n_1)\times(n_0-n_2)}}}
\end{equation}
\end{definition}

\begin{definition}[Ranking error~\citep
{Wang2018}]
\label{def:ranking-error}
Let $f_{IC}(i)$ denote the expected spread of node $i$ in the IC model.
Moreover, let $\phi(k)$ denote the set of top-$k$ nodes that are selected by a specific measure and let $\Phi(k)$ denote
the set of top-$k$ nodes selected by expected spreads ranking of nodes 
within the IC model.
Then, the ranking error $epsilon$ is given as:
\begin{equation} \label{eq:20}
\epsilon := 1-\frac{\sum_{i \in \phi(k)}f_{IC}(i)}{\sum_{j \in \Phi(k)}f_{IC}(i)}
\end{equation}{}
\end{definition}

\begin{definition}[Ranking Monotonicity~\citep
{Wang2018,Liu2016}]
\label{def:monotonicity}
Let $R$ denote a ranking list and $|V|$ the number of nodes in the network.
Moreover, let $|V|_r$ denote the number of nodes with rank $r$ in $R$.
Then, the \emph{monotonicity} of $R$, $M(R)$, is given as:
\begin{equation} \label{eq:21}
M(R) := {\Bigg(1-\frac{\sum_{r \in V}|V|_r \times (|V|_r-1)}{|V| \times (|V|-1)}\Bigg)^2}
\end{equation}
\end{definition}

\newpage

\section{Detailed Experimental Results}
\label{sec:kendall-explore}
\begin{table}[h!]
 \small
\caption{Kendall $\tau$ results for basic measures normalized by $\COD$
}
\centering
\smallskip
\label{table:kendall-explore}
\resizebox*{0.89\textwidth}{0.89 \textheight}{
 \begin{tabular}{||c c c c c c c c c||} 
 \hline
 Data sets & $\COS$ & $\CBuu$ & $\CBuw$ & $\CCdu$ & $\CtCdw$ & $\CEuu$ & $\CKatzdu$ & $wks$ \\ [0.5ex] 
 \hline\hline
Moreno seventh grader & 1.07219 & 0.86828 & 0.0436 & 1.0895 & 1.05326 & 0.43518 & 0.04415 & 0.76777 \\
Moreno dutch college & 0.94924 & 0.48372 & -0.12011 & 1 & 0.99014 & 0.75373 & 0.24781 & 0.74834 \\
Moreno highschool & 1.84211 & 0.59106 & 0.80837 & 1.30892 & 1.74549 & -0.23471 & -0.21467 & 1.07704 \\
Moreno residence hall & 1.16988 & 0.59655 & -0.01666 & 0.87276 & 0.92799 & 0.41392 & 0.20729 & 0.56992 \\
Moreno physicians & 1.5771 & 0.52942 & 0.46768 & 0.40526 & 0.92737 & -0.01444 & 0.05671 & 1.33096 \\
socfb-Haverford76 & 1.28397 & 0.36361 & -0.02064 & 0.38122 & 1.12532 & 0.30771 & -0.32919 & 0.80109 \\
socfb-Simmons81 & 1.22785 & 0.35745 & 0.04244 & 0.27842 & 0.91182 & 0.35444 & -0.26511 & 0.68964 \\
socfb-Swarthmore42 & 1.31375 & 0.399 & -0.0258 & 0.34167 & 1.06816 & 0.31246 & -0.31487 & 0.67667 \\
Petster hamster friendships & 1.85412 & 0.51868 & 0.51424 & -0.36151 & 1.1611 & 0.30978 & 0.09258 & 0.83615 \\
Socfb-Amherst41 & 1.28085 & 0.37866 & -0.04531 & 0.46236 & 1.0534 & 0.30983 & -0.29083 & 0.75061 \\
socfb-Bowdoin47 & 1.28252 & 0.37619 & 0.009 & 0.45425 & 1.07021 & 0.29348 & -0.31617 & 0.55806 \\
socfb-Hamilton46 & 1.27881 & 0.34249 & -0.01253 & 0.46144 & 1.11102 & 0.28117 & -0.34063 & 0.75356 \\
Moreno adolescent health & 1.80561 & 0.56098 & 0.49428 & 1.37365 & 1.93651 & 0.08371 & -0.14765 & 1.36908 \\
socfb-Trinity100 & 1.27496 & 0.33278 & -0.0074 & 0.48476 & 1.05317 & 0.29859 & -0.31929 & 0.787 \\
socfb-USFCA72 & 1.29699 & 0.41294 & 0.03016 & 0.45007 & 0.97061 & 0.36918 & -0.17213 & 0.64924 \\
socfb-Williams40 & 1.25598 & 0.35576 & 0.01324 & 0.52593 & 1.08511 & 0.31502 & -0.28525 & 0.63013 \\
socfb-nips-ego & 0.91424 & 1.06276 & 0.84811 & -0.03844 & 0.36752 & 0.46641 & 0.09795 & 0.83042 \\
socfb-Oberlin44 & 1.35408 & 0.38774 & 0.02021 & 0.40455 & 1.04593 & 0.32949 & -0.28713 & 0.68642 \\
socfb-Smith60 & 1.22492 & 0.38594 & 0.01614 & 0.51633 & 1.00311 & 0.33518 & -0.29024 & 0.75543 \\
socfb-Wellesley22 & 1.37423 & 0.40136 & -0.00929 & 0.34287 & 1.05348 & 0.32625 & -0.28781 & 0.76453 \\
socfb-Vassar85 & 1.31528 & 0.3276 & -0.04416 & 0.46474 & 1.13225 & 0.28187 & -0.35924 & 0.76019 \\
socfb-Middlebury45 & 1.27759 & 0.37768 & 0.00691 & 0.35034 & 1.04371 & 0.32444 & -0.26949 & 0.75653 \\
socfb-Pepperdine86 & 1.30586 & 0.40344 & 0.03167 & 0.47996 & 1.03967 & 0.36903 & -0.20436 & 0.7052 \\
socfb-Colgate88 & 1.27992 & 0.30799 & 0.01587 & 0.42624 & 1.08471 & 0.26193 & -0.36464 & 0.71967 \\
socfb-Santa74 & 1.32182 & 0.36053 & -0.04409 & 0.5468 & 1.10208 & 0.31662 & -0.272 & 0.67879 \\
socfb-Wesleyan43 & 1.32177 & 0.35419 & -0.03332 & 0.48109 & 1.09294 & 0.30385 & -0.31592 & 0.82727 \\
socfb-Mich67 & 1.34677 & 0.44064 & 0.06052 & 0.43828 & 1.0198 & 0.36284 & -0.18038 & 0.63085 \\
socfb-Bucknell39 & 1.28519 & 0.32119 & -0.03325 & 0.60848 & 1.0612 & 0.27474 & -0.37136 & 0.6845 \\
Facebook 2018 tvshow & 1.41447 & 0.47297 & 0.35412 & -0.54842 & -0.11656 & 0.10791 & -0.20833 & 0.73301 \\
socfb-Brandeis99 & 1.26233 & 0.4091 & -0.02065 & 0.60498 & 1.02234 & 0.3764 & -0.1965 & 0.50282 \\
Facebook combined & 1.57412 & 0.3886 & 0.10442 & -0.55989 & 0.23332 & 0.13391 & -0.35056 & 0.6728 \\
socfb-Howard90 & 1.31767 & 0.40883 & -0.01443 & 0.58347 & 1.08569 & 0.38137 & -0.20698 & 0.77934 \\
socfb-Rice31 & 1.28138 & 0.37591 & -0.0007 & 0.4965 & 1.07403 & 0.37021 & -0.23798 & 0.75898 \\
socfb-Rochester38 & 1.28599 & 0.34293 & -0.01326 & 0.52499 & 1.0627 & 0.31673 & -0.25947 & 0.54769 \\
Facebook 2018 politician & 1.39533 & 0.48556 & 0.2948 & -0.70569 & 0.00676 & 0.17049 & -0.19246 & 0.7189 \\
advogato & 0.76081 & 0.7351 & 0.43044 & 0.85957 & 0.91846 & 0.6862 & 0.60825 & 0.47472 \\
Facebook 2018 government & 1.35247 & 0.43016 & 0.15733 & -0.22203 & 0.53554 & 0.18745 & -0.16143 & 0.56916 \\
Wiki-Vote & 1.00791 & 0.66257 & 0.44556 & 0.59118 & 0.60544 & 0.77417 & 0.46799 & 0.55412 \\
Socfb-BC17 & 1.3052 & 0.37091 & -0.03239 & 0.68327 & 1.08855 & 0.37081 & -0.23247 & 0.65282 \\
Facebook 2018 public figure & 1.38879 & 0.57016 & 0.34058 & -0.35101 & 0.18357 & 0.22931 & -0.01495 & 0.65866 \\
socfb-Columbia2 & 1.36419 & 0.45096 & 0.02765 & 0.7161 & 1.09404 & 0.41569 & -0.12648 & 0.52236 \\
Facebook 2018 athletes & 1.1923 & 0.48305 & 0.27088 & -0.31938 & 0.2857 & 0.3161 & -0.09231 & 0.60546 \\
socfb-JMU79 & 1.29742 & 0.36219 & -0.04413 & 0.78601 & 1.07609 & 0.37797 & -0.27027 & 0.33486 \\
Facebook 2018 company & 1.28638 & 0.47791 & 0.34551 & -0.57644 & -0.20005 & 0.20814 & -0.12015 & 0.61745 \\
socfb-UCSB37 & 1.35853 & 0.41187 & -0.01882 & 0.71768 & 1.10934 & 0.39657 & -0.17722 & 0.72479 \\
socfb-UCF52 & 1.37173 & 0.43188 & 0.00187 & 0.67341 & 0.98016 & 0.41325 & -0.14063 & 0.41168 \\
Facebook 2018 new sites & 1.34588 & 0.4677 & 0.20821 & -0.19962 & 0.33479 & 0.28954 & -0.1521 & 0.53644 \\
Deezer RO & 1.30968 & 0.39828 & 0.17782 & -0.79879 & -0.47604 & 0.2095 & -0.28423 & 0.513 \\
Deezer HU & 1.29675 & 0.35853 & 0.0919 & -0.78482 & -0.34168 & 0.26132 & -0.34625 & 0.67605 \\
Deezer HR & 1.37833 & 0.40432 & 0.06735 & -0.1702 & 0.51123 & 0.36759 & -0.22419 & 0.63562 \\
\hline\hline
geometric mean & 1.29864 & 0.43427 & 0.051259 & 0.491424 & 0.725396 & 0.291757 & 0.208434 & 0.67953 \\
 \hline
\end{tabular}
}
\end{table}

\begin{table}[h!]
 \small
\caption{Normalized Kendall $\tau$ results for combined measures}
\centering
\smallskip
\label{table:kendall-combined}
\resizebox*{0.89\textwidth}{0.89 \textheight}{

 \begin{tabular}{||c c c c c c||} 
 \hline
 Data sets & $\SCfa$ & $\SKfa$ & $\SKfb$ & $\SKfc$ & $\CKatzdwrightarrow$  \\ [0.5ex] 
 \hline\hline
Moreno seventh grader & 1.02173 & 1.12895 & 1.07219 & 1.07849 & 1.07219  \\
Moreno dutch college & 0.99964 & 0.042 & 0.94924 & 0.94504 & 0.94924  \\
Moreno highschool & 2.3055 & 1.67887 & 1.86304 & 1.8563 & 1.84658  \\
Moreno residence hall & 1.16266 & 0.21573 & 1.16537 & 1.17091 & 1.16988  \\
Moreno  physicians & 1.90406 & 1.29563 & 1.5627 & 1.57417 & 1.57725  \\
socfb-Haverford76 & 1.29127 & 1.23903 & 1.28214 & 1.31473 & 1.28412  \\
socfb-Simmons81 & 1.26075 & 1.13576 & 1.22459 & 1.25217 & 1.22797  \\
socfb-Swarthmore42 & 1.30656 & 1.26022 & 1.31087 & 1.34036 & 1.31386  \\
Petster hamster friendships & 2.03253 & 1.59719 & 1.83377 & 1.87295 & 1.85556  \\
Socfb-Amherst41 & 1.29013 & 1.23109 & 1.27887 & 1.31191 & 1.28096  \\
socfb-Bowdoin47 & 1.2893 & 1.23733 & 1.28024 & 1.31379 & 1.28264  \\
socfb-Hamilton46 & 1.28768 & 1.2393 & 1.27684 & 1.31151 & 1.27891  \\
Moreno adolescent health & 2.30235 & 0.36232 & 1.78791 & 1.80946 & 1.80672  \\
socfb-Trinity100 & 1.27118 & 1.24274 & 1.27346 & 1.30606 & 1.27505  \\
socfb-USFCA72 & 1.3065 & 1.23325 & 1.29336 & 1.31714 & 1.29715  \\
socfb-Williams40 & 1.25352 & 1.21565 & 1.25386 & 1.28535 & 1.25609  \\
socfb-nips-ego & 1.12072 & 0.81988 & 0.81988 & 0.87383 & 0.96288  \\
socfb-Oberlin44 & 1.36136 & 1.28213 & 1.35199 & 1.38201 & 1.35422  \\
socfb-Smith60 & 1.24548 & 1.17268 & 1.22259 & 1.25609 & 1.22502  \\
socfb-Wellesley22 & 1.36897 & 1.32054 & 1.37127 & 1.39548 & 1.37435  \\
socfb-Vassar85 & 1.31344 & 1.2748 & 1.31266 & 1.34281 & 1.31538  \\
socfb-Middlebury45 & 1.2818 & 1.23765 & 1.27519 & 1.30794 & 1.27773  \\
socfb-Pepperdine86 & 1.29885 & 1.25882 & 1.3012 & 1.33746 & 1.306  \\
socfb-Colgate88 & 1.28686 & 1.24514 & 1.27882 & 1.31266 & 1.28003  \\
socfb-Santa74 & 1.31672 & 1.28173 & 1.31999 & 1.35012 & 1.32193  \\
socfb-Wesleyan43 & 1.32115 & 1.27399 & 1.31886 & 1.34998 & 1.32191  \\
socfb-Mich67 & 1.34482 & 1.26409 & 1.34299 & 1.36394 & 1.34692  \\
socfb-Bucknell39 & 1.28908 & 1.24405 & 1.28313 & 1.31614 & 1.2853  \\
Facebook 2018 tvshow & 1.40682 & 0.8598 & 1.36151 & 1.41537 & 1.42149  \\
socfb-Brandeis99 & 1.26142 & 1.20605 & 1.25883 & 1.29074 & 1.26246  \\
Facebook combined & 1.60939 & 1.30368 & 1.56052 & 1.57443 & 1.57445  \\
socfb-Howard90 & 1.31449 & 1.27997 & 1.31616 & 1.34902 & 1.3178  \\
socfb-Rice31 & 1.28918 & 1.2417 & 1.27935 & 1.31631 & 1.2815  \\
socfb-Rochester38 & 1.29122 & 1.24549 & 1.28376 & 1.31682 & 1.2861  \\
Facebook 2018 politician & 1.48084 & 1.19778 & 1.37895 & 1.40473 & 1.40002  \\
advogato & 1.10156 & 0.67022 & 0.75117 & 0.74454 & 0.94766  \\
Facebook 2018 government & 1.43131 & 1.25692 & 1.34626 & 1.3686 & 1.35283  \\
Wiki-Vote & 1.08119 & 0.99839 & 1.00578 & 0.99949 & 1.00823  \\
Socfb-BC17 & 1.31419 & 1.26612 & 1.30292 & 1.33626 & 1.30531  \\
Facebook 2018 public figure & 1.45592 & 1.1719 & 1.36848 & 1.39434 & 1.39342  \\
socfb-Columbia2 & 1.37154 & 1.30789 & 1.35883 & 1.38909 & 1.36435  \\
Facebook 2018 athletes & 1.32884 & 1.09071 & 1.18449 & 1.20103 & 1.19443  \\
socfb-JMU79 & 1.3225 & 1.25209 & 1.29578 & 1.32809 & 1.29754  \\
Facebook 2018 company & 1.33801 & 0.8942 & 1.24648 & 1.28932 & 1.29482  \\
socfb-UCSB37 & 1.38757 & 1.30484 & 1.35613 & 1.3848 & 1.35867  \\
socfb-UCF52 & 1.38952 & 1.31247 & 1.36919 & 1.389 & 1.37189  \\
Facebook 2018 new sites & 1.46062 & 1.1829 & 1.33249 & 1.35655 & 1.34878  \\
Deezer RO & 1.40734 & 0.80106 & 1.27555 & 1.31152 & 1.31279  \\
Deezer HU & 1.39269 & 1.05544 & 1.28433 & 1.30042 & 1.29783  \\
Deezer HR & 1.48176 & 1.25906 & 1.37196 & 1.38919 & 1.37875  \\
\hline\hline
geometric mean & 1.35576 & 1.04994 & 1.28907 & 1.31493 & 1.30657 \\
 \hline
\end{tabular}
}

\end{table}

\begin{table}[h!]
 \small
\caption{Normalized Kendall $\tau$ results for modifications of gravitational centrality}
\centering
\smallskip
\label{table:kendall-gc}
\resizebox*{0.89\textwidth}{0.89 \textheight}{
 \begin{tabular}{||c c c c c c c||} 
 \hline
 Data sets & $\CGCw$ & $MGC_{ODS}$ & $MGC_{S}$ & $MGC_{SC}$ & $MGC_{SK}$ & $MGC_{wk}$  \\ [0.5ex] 
 \hline\hline
Moreno seventh grader & 0.73459 & 1.16048 & 1.02173 & 1.07849 & 1.10372 & 1.09111 \\
Moreno dutch college & 0.85002 & 1.00804 & 0.99964 & 0.95344 & 0.94924 & 0.91984 \\
Moreno highschool & 1.1495 & 2.31673 & 2.30999 & 1.83834 & 1.84507 & 2.14828 \\
Moreno residence hall & 0.56771 & 1.27052 & 1.16317 & 1.17027 & 1.17066 & 1.20033 \\
Moreno  physicians & 1.52192 & 1.96817 & 1.90406 & 1.57725 & 1.57131 & 1.76854 \\
socfb-Haverford76 & 0.84677 & 1.3151 & 1.2913 & 1.28578 & 1.31109 & 1.34965 \\
socfb-Simmons81 & 0.71641 & 1.27526 & 1.26071 & 1.22916 & 1.25748 & 1.31159 \\
socfb-Swarthmore42 & 0.69794 & 1.33336 & 1.30655 & 1.31462 & 1.33739 & 1.38106 \\
Petster hamster friendships & 0.84828 & 2.04285 & 2.02824 & 1.85425 & 1.87918 & 2.07271 \\
Socfb-Amherst41 & 0.77978 & 1.31293 & 1.29016 & 1.28161 & 1.30967 & 1.34943 \\
socfb-Bowdoin47 & 0.57039 & 1.30414 & 1.28926 & 1.28366 & 1.31301 & 1.3482 \\
socfb-Hamilton46 & 0.78984 & 1.30853 & 1.28771 & 1.27967 & 1.30965 & 1.34618 \\
Moreno adolescent health & 1.6793 & 2.27963 & 2.29936 & 1.80534 & 1.81307 & 2.09881 \\
socfb-Trinity100 & 0.82539 & 1.29106 & 1.27117 & 1.27605 & 1.30344 & 1.32765 \\
socfb-USFCA72 & 0.66755 & 1.32653 & 1.30637 & 1.29776 & 1.31831 & 1.37363 \\
socfb-Williams40 & 0.64841 & 1.27252 & 1.25352 & 1.25718 & 1.28258 & 1.31505 \\
socfb-nips-ego & 1.01414 & 1.25598 & 1.12072 & 0.87383 & 0.87383 & 1.30608 \\
socfb-Oberlin44 & 0.70955 & 1.37339 & 1.36131 & 1.35471 & 1.38163 & 1.4336 \\
socfb-Smith60 & 0.79162 & 1.25405 & 1.24552 & 1.22554 & 1.25862 & 1.30391 \\
socfb-Wellesley22 & 0.79596 & 1.38155 & 1.36873 & 1.37451 & 1.39211 & 1.44012 \\
socfb-Vassar85 & 0.79713 & 1.33655 & 1.31337 & 1.31608 & 1.33868 & 1.38237 \\
socfb-Middlebury45 & 0.79021 & 1.29392 & 1.28173 & 1.27855 & 1.30642 & 1.345 \\
socfb-Pepperdine86 & 0.7316 & 1.31475 & 1.29875 & 1.30663 & 1.33287 & 1.37074 \\
socfb-Colgate88 & 0.74988 & 1.30449 & 1.28689 & 1.28092 & 1.30904 & 1.34626 \\
socfb-Santa74 & 0.70084 & 1.34198 & 1.31664 & 1.32237 & 1.34499 & 1.38677 \\
socfb-Wesleyan43 & 0.87143 & 1.34025 & 1.32113 & 1.32203 & 1.34634 & 1.39005 \\
socfb-Mich67 & 0.64648 & 1.36589 & 1.34467 & 1.34715 & 1.36307 & 1.42902 \\
socfb-Bucknell39 & 0.7113 & 1.3081 & 1.28901 & 1.28633 & 1.31292 & 1.34831 \\
Facebook 2018 tvshow & 0.69339 & 1.45027 & 1.40002 & 1.41356 & 1.41549 & 1.49404 \\
socfb-Brandeis99 & 0.51041 & 1.2783 & 1.26135 & 1.26236 & 1.28888 & 1.33432 \\
Facebook combined & 0.67483 & 1.69784 & 1.60843 & 1.57409 & 1.56488 & 1.67843 \\
socfb-Howard90 & 0.82153 & 1.33585 & 1.31452 & 1.31836 & 1.3412 & 1.39143 \\
socfb-Rice31 & 0.79711 & 1.31074 & 1.28919 & 1.28239 & 1.3122 & 1.36068 \\
socfb-Rochester38 & 0.55915 & 1.30834 & 1.29121 & 1.28625 & 1.31587 & 1.35564 \\
Facebook 2018 politician & 0.71679 & 1.52019 & 1.47809 & 1.39526 & 1.40782 & 1.54248 \\
advogato & 0.47654 & 1.22577 & 1.09978 & 0.76072 & 0.74031 & 1.18128 \\
Facebook 2018 government & 0.57675 & 1.44967 & 1.43062 & 1.3525 & 1.37007 & 1.48547 \\
Wiki-Vote & 0.55593 & 1.08693 & 1.08121 & 1.00789 & 0.9945 & 1.13243 \\
Socfb-BC17 & 0.67427 & 1.32557 & 1.31418 & 1.30554 & 1.33301 & 1.38271 \\
Facebook 2018 public figure & 0.66155 & 1.47636 & 1.45225 & 1.38843 & 1.39538 & 1.53463 \\
socfb-Columbia2 & 0.53067 & 1.37866 & 1.3713 & 1.36421 & 1.38497 & 1.45478 \\
Facebook 2018 athletes & 0.61181 & 1.33842 & 1.32811 & 1.19227 & 1.20419 & 1.35043 \\
socfb-JMU79 & 0.33717 & 1.3333 & 1.32253 & 1.29744 & 1.32671 & 1.38115 \\
Facebook 2018 company & 0.60783 & 1.3633 & 1.33336 & 1.28576 & 1.29077 & 1.41088 \\
socfb-UCSB37 & 0.75196 & 1.39101 & 1.38756 & 1.35906 & 1.38202 & 1.45049 \\
socfb-UCF52 & 0.41494 & 1.4033 & 1.38954 & 1.37175 & 1.3872 & 1.46115 \\
Facebook 2018 new sites & 0.53968 & 1.46971 & 1.45859 & 1.3458 & 1.35829 & 1.50102 \\
Deezer RO & 0.48595 & 1.42253 & 1.40325 & 1.30878 & 1.31349 & 1.42544 \\
Deezer HU & 0.66951 & 1.40481 & 1.39164 & 1.29645 & 1.30283 & 1.40613 \\
Deezer HR & 0.64705 & 1.48671 & 1.48101 & 1.37828 & 1.39161 & 1.49918 \\
\hline\hline
geometric mean & 0.70205 & 1.38455 & 1.35513 & 1.29802 & 1.31421 & 1.40721 \\
 \hline
\end{tabular}
}
\end{table}

\begin{table}[h!]
 \small
\caption{Normalized $\epsilon$ for $\COS$, $wks$, $\CKatzdwrightarrow$, and the proposed combined measures}
\centering
\smallskip
\label{table:error-combined}
\resizebox*{0.89\textwidth}{0.89 \textheight}{
 \begin{tabular}{||c c c c c c c c||}

 \hline
 Data sets & $\COS$ & $wks$ & $\SCfa$ & $\SKfa$ & $\SKfb$ & $\SKfc$ & $\CKatzdwrightarrow$  \\ [0.5ex] 
 \hline\hline
Moreno residence hall & 0.25512 & 0.69927 & 0.29313 & 0.31851 & 0.33243 & 0.33487 & 0.25512 \\
Moreno  physicians & 0.30256 & 0.37274 & 0.30256 & 0.30256 & 0.30256 & 0.30256 & 0.30256 \\
socfb-Haverford76 & 0.26202 & 0.47165 & 0.26202 & 0.22385 & 0.22062 & 0.21282 & 0.26202 \\
socfb-Simmons81 & 0.54007 & 0.71912 & 0.54007 & 0.48921 & 0.49367 & 0.48965 & 0.54007 \\
socfb-Swarthmore42 & 0.46276 & 0.6336 & 0.46276 & 0.35905 & 0.45217 & 0.35905 & 0.46276 \\
Petster hamster friendships & 0.39754 & 0.61945 & 0.39754 & 0.35454 & 0.36399 & 0.3498 & 0.39754 \\
Socfb-Amherst41 & 0.36125 & 0.68422 & 0.36125 & 0.31794 & 0.34848 & 0.28005 & 0.36125 \\
socfb-Bowdoin47 & 0.30648 & 0.50406 & 0.30648 & 0.25929 & 0.25929 & 0.24939 & 0.30648 \\
socfb-Hamilton46 & 0.37336 & 0.78778 & 0.37336 & 0.25364 & 0.30259 & 0.25686 & 0.37336 \\
Moreno adolescent health & 0.42094 & 0.47341 & 0.42094 & 0.42358 & 0.42094 & 0.4113 & 0.42094 \\
socfb-Trinity100 & 0.46712 & 0.65428 & 0.46712 & 0.36999 & 0.38724 & 0.35247 & 0.46712 \\
socfb-USFCA72 & 0.38633 & 0.84379 & 0.38633 & 0.3566 & 0.36694 & 0.35349 & 0.38633 \\
socfb-Williams40 & 0.49189 & 0.78705 & 0.49189 & 0.41761 & 0.41761 & 0.41284 & 0.49189 \\
socfb-nips-ego & 1 & 1 & 1 & 1 & 1 & 1 & 1 \\
socfb-Oberlin44 & 0.3785 & 0.47519 & 0.3785 & 0.31897 & 0.35904 & 0.31644 & 0.3785 \\
socfb-Smith60 & 0.24135 & 0.41943 & 0.24135 & 0.2032 & 0.2032 & 0.2032 & 0.24135 \\
socfb-Wellesley22 & 0.48079 & 0.76193 & 0.48079 & 0.44613 & 0.46375 & 0.41113 & 0.48079 \\
socfb-Vassar85 & 0.25668 & 0.62991 & 0.25668 & 0.25211 & 0.25211 & 0.21918 & 0.25668 \\
socfb-Middlebury45 & 0.46514 & 0.57362 & 0.46514 & 0.38526 & 0.38526 & 0.31756 & 0.46514 \\
socfb-Pepperdine86 & 0.3885 & 0.75458 & 0.3885 & 0.3325 & 0.35018 & 0.31855 & 0.3885 \\
socfb-Colgate88 & 0.51404 & 0.72975 & 0.51404 & 0.47323 & 0.49853 & 0.45644 & 0.51404 \\
socfb-Santa74 & 0.32979 & 0.59999 & 0.32979 & 0.27099 & 0.30347 & 0.24535 & 0.32979 \\
socfb-Wesleyan43 & 0.29342 & 0.58256 & 0.29342 & 0.25977 & 0.25977 & 0.26154 & 0.29342 \\
socfb-Mich67 & 0.67358 & 0.83665 & 0.67358 & 0.51844 & 0.52565 & 0.48376 & 0.67358 \\
socfb-Bucknell39 & 0.41871 & 0.59615 & 0.41871 & 0.38348 & 0.39553 & 0.35257 & 0.41871 \\
Facebook 2018 tvshow & 0.61928 & 0.59371 & 0.56083 & 0.56107 & 0.56107 & 0.56752 & 0.61239 \\
socfb-Brandeis99 & 0.32414 & 0.53625 & 0.32414 & 0.26547 & 0.26547 & 0.24321 & 0.32414 \\
Facebook combined & 0.55791 & 0.66992 & 0.56234 & 0.53208 & 0.53208 & 0.51662 & 0.55791 \\
socfb-Howard90 & 0.40548 & 0.71537 & 0.40548 & 0.27983 & 0.33532 & 0.26495 & 0.40548 \\
socfb-Rice31 & 0.44395 & 0.57937 & 0.44395 & 0.39381 & 0.39684 & 0.39381 & 0.44395 \\
socfb-Rochester38 & 0.27909 & 0.60616 & 0.27909 & 0.23 & 0.23492 & 0.22546 & 0.27909 \\
Facebook 2018 politician & 0.53271 & 0.67737 & 0.53271 & 0.50283 & 0.50283 & 0.51869 & 0.53271 \\
advogato & 1.04794 & 0.73164 & 1.04794 & 0.98956 & 0.95241 & 0.98956 & 1.03962 \\
Facebook 2018 government & 0.48642 & 0.65609 & 0.48642 & 0.4455 & 0.4455 & 0.4455 & 0.48642 \\
Wiki-Vote & 0.47074 & 0.44612 & 0.47074 & 0.53308 & 0.56687 & 0.68054 & 0.47074 \\
Socfb-BC17 & 0.57718 & 0.70532 & 0.57718 & 0.47559 & 0.56096 & 0.47666 & 0.57718 \\
Facebook 2018 public figure & 0.66701 & 0.66535 & 0.66701 & 0.62448 & 0.63913 & 0.62449 & 0.67371 \\
socfb-Columbia2 & 0.66694 & 0.81704 & 0.66694 & 0.56962 & 0.56962 & 0.56962 & 0.66694 \\
Facebook 2018 athletes & 0.92226 & 0.88353 & 0.92226 & 0.83905 & 0.8565 & 0.82413 & 0.92226 \\
socfb-JMU79 & 0.53781 & 0.81163 & 0.53781 & 0.43576 & 0.43576 & 0.43385 & 0.53781 \\
Facebook 2018 company & 0.76854 & 0.70986 & 0.75373 & 0.72783 & 0.72783 & 0.72783 & 0.76854 \\
socfb-UCSB37 & 0.66844 & 0.73694 & 0.66844 & 0.62428 & 0.65571 & 0.61348 & 0.66844 \\
socfb-UCF52 & 0.52686 & 0.72915 & 0.52686 & 0.50551 & 0.52086 & 0.49955 & 0.52686 \\
Facebook 2018 new sites & 0.70388 & 0.71843 & 0.70388 & 0.65429 & 0.67445 & 0.65429 & 0.70388 \\
Deezer RO & 0.73214 & 0.82637 & 0.73214 & 0.73214 & 0.73214 & 0.70391 & 0.73214 \\
Deezer HU & 0.6953 & 0.70415 & 0.6953 & 0.68203 & 0.68695 & 0.67093 & 0.6953 \\
Deezer HR & 0.75138 & 0.81517 & 0.75138 & 0.69428 & 0.72058 & 0.69603 & 0.75138 \\
\hline\hline
geometric mean & 0.47425 & 0.65788 & 0.47453 & 0.42411 & 0.43921 & 0.41501 & 0.47415 \\
 \hline
\end{tabular}
}
\end{table}

\begin{table}[h!]
 \small
\caption{Normalized $\epsilon$ for modifications of gravitational centrality.}
\centering
\smallskip
\label{table:error-gc}
\resizebox*{0.89\textwidth}{0.89 \textheight}{
 \begin{tabular}{||c c c c c c c||} 
 \hline
 Data sets & $\CGCw$ & $MGC_{ODS}$ & $MGC_{S}$ & $MGC_{SC}$ & $MGC_{SK}$ & $MGC_{wk}$  \\ [0.5ex] 
 \hline\hline
Moreno residence hall & 2.448 & 0.11315 & 0.42324 & 0.42324 & 0.51414 & 0.38118 \\
Moreno  physicians & 0.20283 & 0.10236 & 0.21894 & 0.21894 & 0.18107 & 0.2643 \\
socfb-Haverford76 & 0.11135 & 0.01677 & 0.01707 & 0.01707 & 0.02279 & 0.08911 \\
socfb-Simmons81 & 0.32375 & 0.0525 & 0.05035 & 0.05035 & 0.05267 & 0.22209 \\
socfb-Swarthmore42 & 0.09462 & 0.02152 & 0.04461 & 0.04461 & 0.0505 & 0.09746 \\
Petster hamster friendships & 0.43121 & 0.11701 & 0.15453 & 0.14662 & 0.18088 & 0.32448 \\
Socfb-Amherst41 & 0.06586 & 0.02541 & 0.02802 & 0.02802 & 0.02802 & 0.09175 \\
socfb-Bowdoin47 & 0.20356 & 0.01704 & 0.02473 & 0.02473 & 0.03083 & 0.08075 \\
socfb-Hamilton46 & 0.06337 & 0.03676 & 0.03763 & 0.03763 & 0.04067 & 0.13756 \\
Moreno adolescent health & 0.26861 & 0.18761 & 0.13217 & 0.13217 & 0.13217 & 0.3275 \\
socfb-Trinity100 & 0.08626 & 0.04673 & 0.04301 & 0.04301 & 0.04422 & 0.15279 \\
socfb-USFCA72 & 0.14782 & 0.05549 & 0.08039 & 0.08039 & 0.08039 & 0.23444 \\
socfb-Williams40 & 0.18012 & 0.04529 & 0.05901 & 0.05901 & 0.05901 & 0.19405 \\
socfb-nips-ego & 1 & 1 & 1 & 1 & 1 & 1 \\
socfb-Oberlin44 & 0.10535 & 0.10038 & 0.06695 & 0.06695 & 0.05149 & 0.12517 \\
socfb-Smith60 & 0.1146 & 0.03551 & 0.0295 & 0.0295 & 0.03189 & 0.16791 \\
socfb-Wellesley22 & 0.17402 & 0.01778 & 0.05613 & 0.05613 & 0.05921 & 0.22957 \\
socfb-Vassar85 & 0.03984 & 0.02785 & 0.03876 & 0.03876 & 0.03876 & 0.07046 \\
socfb-Middlebury45 & 0.10623 & 0.03542 & 0.02143 & 0.02143 & 0.02863 & 0.18416 \\
socfb-Pepperdine86 & 0.10288 & 0.03321 & 0.03023 & 0.03023 & 0.02289 & 0.11396 \\
socfb-Colgate88 & 0.23763 & 0.07742 & 0.03966 & 0.03966 & 0.03966 & 0.29269 \\
socfb-Santa74 & 0.08188 & 0.04581 & 0.03994 & 0.03994 & 0.03751 & 0.12454 \\
socfb-Wesleyan43 & 0.09717 & 0.0468 & 0.03625 & 0.03625 & 0.03142 & 0.13584 \\
socfb-Mich67 & 0.08518 & 0.09433 & 0.10884 & 0.10884 & 0.10278 & 0.29759 \\
socfb-Bucknell39 & 0.19714 & 0.04229 & 0.04841 & 0.04841 & 0.04841 & 0.18054 \\
Facebook 2018 tvshow & 0.18434 & 0.17221 & 0.24594 & 0.26484 & 0.26063 & 0.42753 \\
socfb-Brandeis99 & 0.17139 & 0.0196 & 0.01558 & 0.01558 & 0.01558 & 0.11424 \\
Facebook combined & 0.49939 & 0.19561 & 0.12713 & 0.12713 & 0.12713 & 0.16093 \\
socfb-Howard90 & 0.0792 & 0.05839 & 0.06017 & 0.06017 & 0.04334 & 0.11557 \\
socfb-Rice31 & 0.072 & 0.0766 & 0.06589 & 0.06589 & 0.06419 & 0.14502 \\
socfb-Rochester38 & 0.10521 & 0.03811 & 0.04021 & 0.04021 & 0.03391 & 0.1071 \\
Facebook 2018 politician & 0.15745 & 0.09422 & 0.19236 & 0.19236 & 0.20892 & 0.44732 \\
advogato & 0.27719 & 0.33991 & 0.56697 & 0.56697 & 0.59685 & 0.8474 \\
Facebook 2018 government & 0.13169 & 0.13371 & 0.18233 & 0.18233 & 0.15915 & 0.31694 \\
Wiki-Vote & 0.10855 & 0.49188 & 0.65787 & 0.65787 & 1.32246 & 1.58915 \\
Socfb-BC17 & 0.08558 & 0.06673 & 0.07063 & 0.07063 & 0.04905 & 0.21086 \\
Facebook 2018 public figure & 0.21784 & 0.17261 & 0.31177 & 0.2801 & 0.231 & 0.54453 \\
socfb-Columbia2 & 0.16332 & 0.12478 & 0.10754 & 0.10754 & 0.10754 & 0.26202 \\
Facebook 2018 athletes & 0.22149 & 0.31631 & 0.22155 & 0.22155 & 0.17837 & 0.70301 \\
socfb-JMU79 & 0.28261 & 0.15332 & 0.13874 & 0.13874 & 0.13874 & 0.27724 \\
Facebook 2018 company & 0.2098 & 0.29788 & 0.22984 & 0.22079 & 0.21741 & 0.62358 \\
socfb-UCSB37 & 0.14998 & 0.09283 & 0.06006 & 0.06006 & 0.05773 & 0.19566 \\
socfb-UCF52 & 0.13389 & 0.13706 & 0.16518 & 0.16518 & 0.12535 & 0.3586 \\
Facebook 2018 new sites & 0.31945 & 0.23881 & 0.23164 & 0.23164 & 0.20128 & 0.53716 \\
Deezer RO & 0.30249 & 0.26882 & 0.27234 & 0.27234 & 0.25782 & 0.51047 \\
Deezer HU & 0.20487 & 0.19902 & 0.21211 & 0.21211 & 0.21211 & 0.45621 \\
Deezer HR & 0.26697 & 0.33505 & 0.26885 & 0.26885 & 0.26257 & 0.53277 \\
\hline\hline
geometric mean & 0.16681 & 0.08432 & 0.09198 & 0.09174 & 0.09066 & 0.24280 \\
 \hline
\end{tabular}
}
\end{table}

\begin{table}[h!]
 \small
\caption{Ranking Monotonicity values for $\COS$, $wks$, $\CKatzdwrightarrow$, and the proposed combined measures}
\centering
\smallskip
\label{table:monotonicity-combined}
\resizebox*{0.89\textwidth}{0.89 \textheight}{
 \begin{tabular}{||c c c c c c c c||} 
 \hline
 Data sets & $\COS$ & $wks$ & $\SCfa$ & $\SKfa$ & $\SKfb$ & $\SKfc$ & $\CKatzdwrightarrow$  \\ [0.5ex] 
 \hline\hline
Moreno seventh grader & 1 & 0.0398 & 1 & 1 & 1 & 1 & 1 \\
Moreno dutch college & 1 & 0.01513 & 1 & 1 & 1 & 1 & 1 \\
Moreno highschool & 0.99504 & 0.00714 & 1 & 1 & 1 & 1 & 1 \\
Moreno residence hall & 1 & 0.00208 & 1 & 1 & 1 & 1 & 1 \\
Moreno  physicians & 0.99806 & 0.04491 & 1 & 1 & 1 & 1 & 1 \\
socfb-Haverford76 & 0.99989 & 0.01022 & 0.99993 & 1 & 1 & 1 & 0.99993 \\
socfb-Simmons81 & 0.99944 & 0.02382 & 0.9999 & 0.99998 & 0.99998 & 0.99998 & 0.9999 \\
socfb-Swarthmore42 & 0.99985 & 0.01822 & 0.99997 & 1 & 1 & 1 & 0.99997 \\
Petster hamster friendships & 0.84912 & 0.02653 & 0.91699 & 0.9191 & 0.91994 & 0.91994 & 0.91699 \\
Socfb-Amherst41 & 0.99981 & 0.01556 & 0.99995 & 1 & 1 & 1 & 0.99995 \\
socfb-Bowdoin47 & 0.9998 & 0.01767 & 0.99998 & 1 & 1 & 1 & 0.99998 \\
socfb-Hamilton46 & 0.99989 & 0.0134 & 0.99997 & 1 & 1 & 1 & 0.99997 \\
Moreno adolescent health & 0.91861 & 0.03401 & 0.99995 & 0.99995 & 0.99995 & 0.99995 & 0.99998 \\
socfb-Trinity100 & 0.99991 & 0.01714 & 0.99999 & 1 & 1 & 1 & 0.99999 \\
socfb-USFCA72 & 0.99824 & 0.03555 & 0.99986 & 0.99998 & 0.99998 & 0.99998 & 0.99986 \\
socfb-Williams40 & 0.99981 & 0.01731 & 0.99997 & 1 & 1 & 1 & 0.99997 \\
socfb-nips-ego & 0.99997 & 0.99993 & 0.99998 & 0.99999 & 0.99999 & 0.99999 & 0.99998 \\
socfb-Oberlin44 & 0.99943 & 0.02982 & 0.99994 & 0.99999 & 0.99999 & 0.99999 & 0.99994 \\
socfb-Smith60 & 0.99944 & 0.02601 & 0.99984 & 1 & 1 & 1 & 0.99984 \\
socfb-Wellesley22 & 0.99953 & 0.02742 & 0.99993 & 1 & 1 & 1 & 0.99993 \\
socfb-Vassar85 & 0.99985 & 0.01426 & 0.99996 & 1 & 1 & 1 & 0.99997 \\
socfb-Middlebury45 & 0.99946 & 0.02197 & 0.99995 & 1 & 1 & 1 & 0.99995 \\
socfb-Pepperdine86 & 0.99932 & 0.02937 & 0.99985 & 0.99995 & 0.99995 & 0.99995 & 0.99986 \\
socfb-Colgate88 & 0.99985 & 0.0142 & 0.99999 & 1 & 1 & 1 & 0.99999 \\
socfb-Santa74 & 0.99973 & 0.0202 & 0.9999 & 0.99999 & 0.99999 & 0.99999 & 0.9999 \\
socfb-Wesleyan43 & 0.99966 & 0.02337 & 0.99996 & 1 & 1 & 1 & 0.99996 \\
socfb-Mich67 & 0.99711 & 0.04927 & 0.99971 & 0.99994 & 0.99994 & 0.99994 & 0.99971 \\
socfb-Bucknell39 & 0.99986 & 0.01962 & 0.99997 & 1 & 1 & 1 & 0.99997 \\
Facebook 2018 tvshow & 0.82555 & 0.24428 & 0.96548 & 0.96886 & 0.96911 & 0.96911 & 0.96936 \\
socfb-Brandeis99 & 0.99935 & 0.03458 & 0.99991 & 0.99998 & 0.99998 & 0.99998 & 0.99991 \\
Facebook combined & 0.99051 & 0.06099 & 0.99831 & 0.99998 & 0.99998 & 0.99998 & 0.99845 \\
socfb-Howard90 & 0.99954 & 0.0265 & 0.99995 & 1 & 1 & 1 & 0.99995 \\
socfb-Rice31 & 0.99966 & 0.02558 & 0.99994 & 1 & 1 & 1 & 0.99994 \\
socfb-Rochester38 & 0.99954 & 0.03071 & 0.99994 & 1 & 1 & 1 & 0.99994 \\
Facebook 2018 politician & 0.9271 & 0.17586 & 0.9884 & 0.99181 & 0.99185 & 0.99185 & 0.98835 \\
advogato & 0.98435 & 0.43515 & 0.99883 & 0.99825 & 0.99825 & 0.99825 & 0.99661 \\
Facebook 2018 government & 0.97943 & 0.11658 & 0.9968 & 0.99895 & 0.99895 & 0.99895 & 0.99686 \\
Wiki-Vote & 0.9655 & 0.40037 & 0.98887 & 0.96793 & 0.96793 & 0.96793 & 0.98887 \\
Socfb-BC17 & 0.9994 & 0.07503 & 0.99998 & 0.99998 & 0.99998 & 0.99998 & 0.99998 \\
Facebook 2018 public figure & 0.89544 & 0.29705 & 0.98941 & 0.9802 & 0.9803 & 0.9802 & 0.99001 \\
socfb-Columbia2 & 0.99774 & 0.11542 & 0.99983 & 0.99977 & 0.99977 & 0.99977 & 0.99983 \\
Facebook 2018 athletes & 0.94458 & 0.24277 & 0.99424 & 0.99567 & 0.9957 & 0.9957 & 0.99421 \\
socfb-JMU79 & 0.9993 & 0.08415 & 0.99998 & 0.99999 & 0.99999 & 0.99999 & 0.99998 \\
Facebook 2018 company & 0.82812 & 0.32434 & 0.97689 & 0.96469 & 0.96491 & 0.96499 & 0.97941 \\
socfb-UCSB37 & 0.99859 & 0.11029 & 0.99996 & 0.99994 & 0.99994 & 0.99994 & 0.99996 \\
socfb-UCF52 & 0.99739 & 0.11593 & 0.99989 & 0.99995 & 0.99995 & 0.99995 & 0.99989 \\
Facebook 2018 new sites & 0.94966 & 0.31787 & 0.99659 & 0.99441 & 0.9944 & 0.99441 & 0.99685 \\
Deezer RO & 0.77783 & 0.46095 & 0.99208 & 0.95579 & 0.95586 & 0.9559 & 0.9949 \\
Deezer HU & 0.87108 & 0.39118 & 0.99695 & 0.98997 & 0.99008 & 0.99004 & 0.99765 \\
Deezer HR & 0.9742 & 0.38287 & 0.99946 & 0.99833 & 0.99835 & 0.99834 & 0.99949 \\
\hline\hline
geometric mean & 0.97148 & 0.05049 & 0.99585 & 0.99435 & 0.99438 & 0.99438 & 0.99603 \\
\hline
\end{tabular}
}
\end{table}

\begin{table}[h!]
\small
\caption{Ranking Monotonicity values for modifications of gravitational centrality.}
\centering
\smallskip
\label{table:monotonicity-gc}
\resizebox*{0.89\textwidth}{0.89 \textheight}{
 \begin{tabular}{||c c c c c c c||} 
 \hline
 Data sets & $\CGCw$ & $MGC_{ODS}$ & $MGC_{S}$ & $MGC_{SC}$ & $MGC_{SK}$ & $MGC_{wk}$  \\ [0.5ex] 
 \hline\hline
Moreno seventh grader & 0.18579 & 1 & 1 & 1 & 1 & 1 \\
Moreno dutch college & 0.47822 & 1 & 1 & 1 & 1 & 1 \\
Moreno highschool & 0.08498 & 1 & 1 & 1 & 1 & 1 \\
Moreno residence hall & 0.02816 & 1 & 1 & 1 & 1 & 1 \\
Moreno  physicians & 0.73685 & 0.99993 & 0.99993 & 0.99993 & 1 & 1 \\
socfb-Haverford76 & 0.23123 & 0.99956 & 0.99956 & 0.99956 & 0.99977 & 0.99993 \\
socfb-Simmons81 & 0.17236 & 0.99903 & 0.99903 & 0.99903 & 0.99908 & 0.9999 \\
socfb-Swarthmore42 & 0.12773 & 0.99953 & 0.99953 & 0.99953 & 0.99956 & 0.99997 \\
Petster hamster friendships & 0.11237 & 0.91399 & 0.91356 & 0.91442 & 0.94663 & 0.91741 \\
Socfb-Amherst41 & 0.18578 & 0.99965 & 0.99965 & 0.99965 & 0.99986 & 0.99995 \\
socfb-Bowdoin47 & 0.08701 & 0.99968 & 0.99968 & 0.99968 & 0.99978 & 0.99998 \\
socfb-Hamilton46 & 0.18893 & 0.99976 & 0.99976 & 0.99976 & 0.99989 & 0.99997 \\
Moreno adolescent health & 0.75033 & 0.99994 & 0.99994 & 0.99994 & 0.99994 & 1 \\
socfb-Trinity100 & 0.23114 & 0.99988 & 0.99988 & 0.99988 & 0.99993 & 0.99999 \\
socfb-USFCA72 & 0.14526 & 0.99926 & 0.99926 & 0.99926 & 0.9995 & 0.99986 \\
socfb-Williams40 & 0.11667 & 0.99985 & 0.99985 & 0.99985 & 0.99991 & 0.99997 \\
socfb-nips-ego & 0.99989 & 0.99996 & 0.99996 & 0.99996 & 0.99997 & 0.99998 \\
socfb-Oberlin44 & 0.15114 & 0.99958 & 0.99958 & 0.99958 & 0.99963 & 0.99994 \\
socfb-Smith60 & 0.23876 & 0.99928 & 0.99928 & 0.99928 & 0.99957 & 0.99984 \\
socfb-Wellesley22 & 0.18772 & 0.99947 & 0.99947 & 0.99947 & 0.99957 & 0.99993 \\
socfb-Vassar85 & 0.1906 & 0.99979 & 0.99979 & 0.99979 & 0.99987 & 0.99997 \\
socfb-Middlebury45 & 0.20712 & 0.99973 & 0.99973 & 0.99973 & 0.99985 & 0.99995 \\
socfb-Pepperdine86 & 0.17224 & 0.99952 & 0.99952 & 0.99952 & 0.99976 & 0.99986 \\
socfb-Colgate88 & 0.16582 & 0.9998 & 0.9998 & 0.9998 & 0.99986 & 0.99999 \\
socfb-Santa74 & 0.1379 & 0.99968 & 0.99968 & 0.99968 & 0.99985 & 0.9999 \\
socfb-Wesleyan43 & 0.26978 & 0.99973 & 0.99973 & 0.99973 & 0.99984 & 0.99996 \\
socfb-Mich67 & 0.14673 & 0.99873 & 0.99873 & 0.99873 & 0.99933 & 0.99971 \\
socfb-Bucknell39 & 0.15732 & 0.99981 & 0.99981 & 0.99981 & 0.99987 & 0.99997 \\
Facebook 2018 tvshow & 0.36418 & 0.96179 & 0.96123 & 0.96304 & 0.97882 & 0.97219 \\
socfb-Brandeis99 & 0.09087 & 0.99956 & 0.99956 & 0.99956 & 0.99974 & 0.99991 \\
Facebook combined & 0.14512 & 0.99672 & 0.99672 & 0.99672 & 0.99785 & 0.99848 \\
socfb-Howard90 & 0.21934 & 0.9997 & 0.9997 & 0.9997 & 0.99981 & 0.99995 \\
socfb-Rice31 & 0.21706 & 0.99964 & 0.99964 & 0.99964 & 0.99979 & 0.99994 \\
socfb-Rochester38 & 0.10144 & 0.99974 & 0.99974 & 0.99974 & 0.99987 & 0.99994 \\
Facebook 2018 politician & 0.28619 & 0.98412 & 0.9843 & 0.98447 & 0.99379 & 0.98926 \\
advogato & 0.48197 & 0.99913 & 0.99913 & 0.99917 & 0.99945 & 0.99952 \\
Facebook 2018 government & 0.18878 & 0.99482 & 0.99482 & 0.99482 & 0.9976 & 0.99693 \\
Wiki-Vote & 0.49329 & 0.98226 & 0.98226 & 0.98226 & 0.98252 & 0.98928 \\
Socfb-BC17 & 0.21013 & 0.99991 & 0.99991 & 0.99991 & 0.99995 & 0.99998 \\
Facebook 2018 public figure & 0.39146 & 0.98715 & 0.98696 & 0.9874 & 0.99447 & 0.99072 \\
socfb-Columbia2 & 0.18284 & 0.99967 & 0.99967 & 0.99967 & 0.99979 & 0.99983 \\
Facebook 2018 athletes & 0.346 & 0.9918 & 0.99173 & 0.99186 & 0.99642 & 0.9945 \\
socfb-JMU79 & 0.11089 & 0.9999 & 0.9999 & 0.9999 & 0.99994 & 0.99998 \\
Facebook 2018 company & 0.4071 & 0.9757 & 0.97492 & 0.97652 & 0.98878 & 0.98193 \\
socfb-UCSB37 & 0.27912 & 0.99987 & 0.99987 & 0.99987 & 0.99993 & 0.99996 \\
socfb-UCF52 & 0.15485 & 0.99969 & 0.99969 & 0.99969 & 0.99983 & 0.99989 \\
Facebook 2018 new sites & 0.37889 & 0.99599 & 0.99589 & 0.99612 & 0.99836 & 0.99729 \\
Deezer RO & 0.50332 & 0.99347 & 0.99286 & 0.99403 & 0.99781 & 0.99563 \\
Deezer HU & 0.49947 & 0.9968 & 0.99666 & 0.9969 & 0.99836 & 0.99786 \\
Deezer HR & 0.4801 & 0.99913 & 0.99911 & 0.99914 & 0.99943 & 0.99951 \\
\hline\hline
geometric mean & 0.21740 & 0.99513 & 0.99508 & 0.99521 & 0.99722 & 0.99628 \\
 \hline
\end{tabular}
}
\end{table}

\label{sec:app-exp-data}

\end{document}